\documentclass[pdflatex,sn-mathphys-num]{sn-jnl}

\usepackage{subcaption}
\usepackage{rotating} 
\usepackage{array}    
\usepackage{eufrak}
\usepackage{graphicx}%
\usepackage{multirow}%
\usepackage{amsmath,amssymb,amsfonts}%
\usepackage{amsthm}%
\usepackage{mathrsfs}%
\usepackage[title]{appendix}%
\usepackage[dvipsnames]{xcolor}%
\usepackage{textcomp}%
\usepackage{manyfoot}%
\usepackage{booktabs}%
\usepackage{algorithm}%
\usepackage{algorithmicx}%
\usepackage{algpseudocode}%
\usepackage{listings}%
\usepackage{multirow}

\theoremstyle{thmstyleone}%
\theoremstyle{thmstyletwo}%

\theoremstyle{thmstylethree}%
\newtheorem{definition}{Definition}%

\begin{document}

\title[Optimizing Overlap in Tree-Based Indexing Structures for Enhanced $k$-NN Search Efficiency]{Optimizing Overlap in Tree-Based Indexing Structures for Enhanced $k$-NN Search Efficiency}



\author*[1,3]{\fnm{Ala-Eddine} \sur{Benrazek}}\email{alaeddine.benrazek@univ-djelfa.dz}

\author[2,3]{\fnm{Zineddin} \sur{Kouahla}}\email{kouahla.zineddine@univ-guelma.dz}

\author[2,3]{\fnm{Brahim} \sur{Farou}}\email{farou.brahim@univ-guelma.dz}

\author[2,3]{\fnm{Hamid} \sur{Seridi}}\email{seridi.hamid@univ-guelma.dz}
\author[2]{\fnm{Ibtissem} \sur{Kemouguette}}\email{kemouguetteibtissem@gmail.com}

\affil[1]{\orgdiv{Department of Computer Science}, \orgname{Ziane Achour University}, \orgaddress{\city{Djelfa}, \postcode{17000}, \country{Algeria}}}

\affil[2]{\orgdiv{Department of Computer Science}, \orgname{8 mai 1945 University}, \orgaddress{\city{Guelma}, \postcode{24000}, \country{Algeria}}}

\affil[3]{\orgdiv{Labstic Laboratory}, \orgname{8 mai 1945 University}, \orgaddress{\city{Guelma}, \postcode{24000}, \country{Algeria}}}


\abstract{The proliferation of interconnected devices in the Internet of Things (IoT) has led to an exponential increase in data, commonly known as Big IoT Data. Efficient retrieval of this heterogeneous data demands a robust indexing mechanism for effective organization. However, a significant challenge remains: the overlap in data space partitions during index construction. This overlap increases node access during search and retrieval, resulting in higher resource consumption, performance bottlenecks, and impedes system scalability. To address this issue, we propose three innovative heuristics designed to quantify and strategically reduce data space partition overlap. The volume-based method (VBM) offers a detailed assessment by calculating the intersection volume between partitions, providing deeper insights into spatial relationships. The distance-based method (DBM) enhances efficiency by using the distance between partition centers and radii to evaluate overlap, offering a streamlined yet accurate approach. Finally, the object-based method (OBM) provides a practical solution by counting objects across multiple partitions, delivering an intuitive understanding of data space dynamics. Experimental results demonstrate the effectiveness of these methods in reducing search time, underscoring their potential to improve data space partitioning and enhance overall system performance.}

\keywords{Data Indexing, Similarity Search, $k$-NN Search, Data Partitioning, Overlap Optimization, Tree-Based Indexing.}



\maketitle

\section{Introduction}\label{sec1}

\begin{figure}[h]
    \centering
    \includegraphics[width=0.8\linewidth]{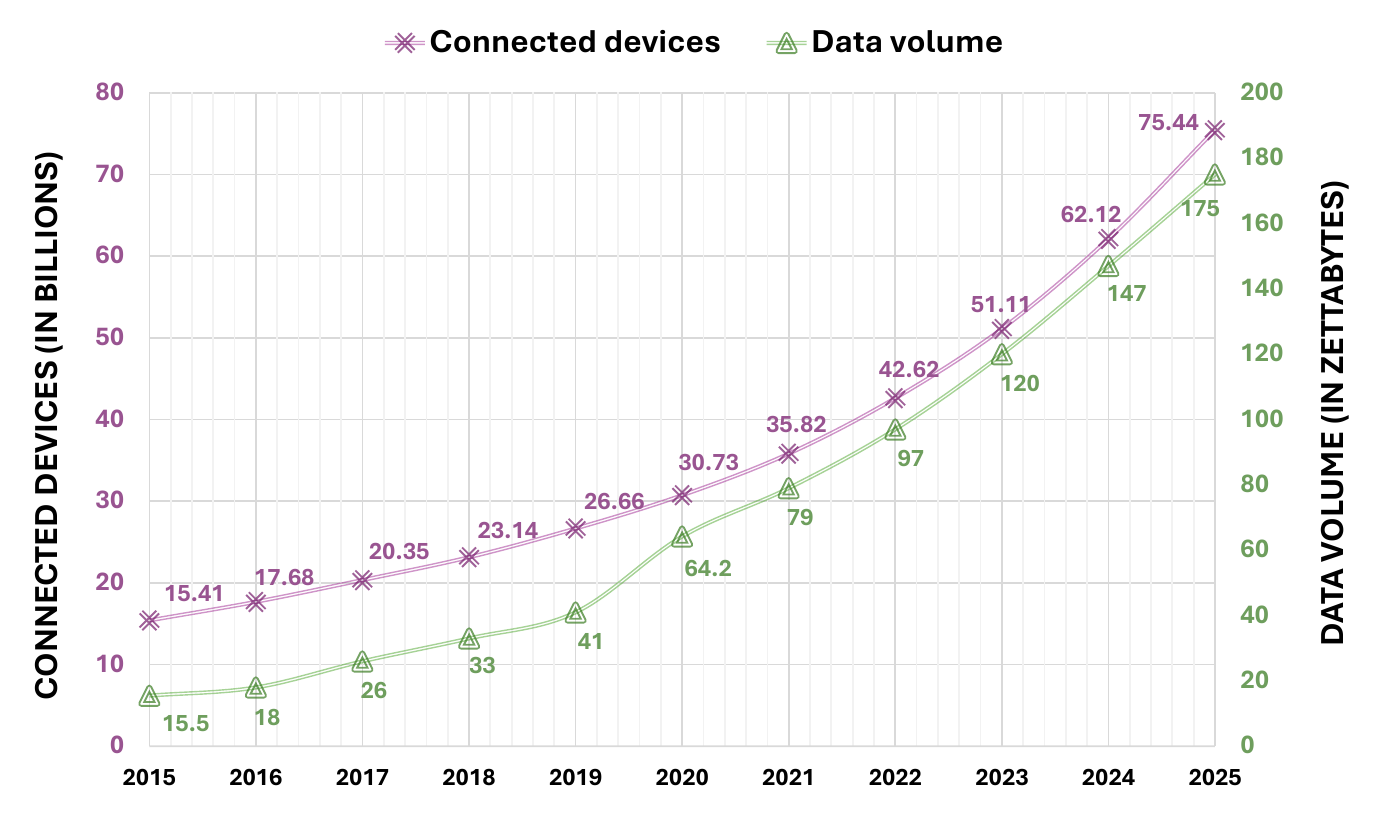}
    \caption{Growth of Connected Devices and Data Volume (2015-2025).}
    \label{fig:Statistics}
\end{figure}

The Internet of Things (IoT) has rapidly emerged as one of the most transformative technologies of the 21st century. This paradigm is characterized by intelligent, self-configuring devices capable of communicating with one another through a global network infrastructure \cite{ge2018big}. Since Kevin Ashton introduced the concept of IoT in 1999 \cite{ashton2009internet}, the number of connected devices has skyrocketed from 8 billion in 2018 to an estimated 75 billion by 2025 \cite{statista2016internet}. Alongside this growth, the volume of generated data has increased from 33 ZB in 2018 to a projected 175 ZB by 2025 \cite{rydning2018digitization}, marking the dawn of the big IoT data era (see Figure \ref{fig:Statistics}).

This exponential growth of data in IoT-based applications offers significant opportunities for value creation but also highlights the urgent for efficient data storage and retrieval mechanisms \cite{benrazek2020efficient}. Similarity search-based data retrieval is widely adopted for its ability to retrieve complex data based on intrinsic characteristics \cite{shimomura2021survey}. However, its effectiveness is often limited by high computational costs, especially in big data settings. To effectively leverage this technique and avoid the brute force sequential scan, robust indexing methods are essential for organizing large datasets in a manner that facilitates the efficient retrieval of similar data \cite{chen2022indexing, kouahla2021survey}. In this context, the concept of metric space plays a crucial role, allowing objects to be mapped into a domain where they can be compared using a distance function \cite{jasbick2020some}. Typically, indexing techniques within this space divide data into hierarchical regions centered around selected representative objects, known as pivots, each with a defined covering radius. While each region is associated with a pivot, every object within the region must be explicitly assigned.

Several metric index methods, such as the M-tree \cite{ciaccia1997m}, Slim-Tree \cite{traina2000slim}, and more recently, the BCCF-tree \cite{benrazek2020efficient} and its extension, the B3CF-tree \cite{khettabi2022clustering}, have been developed to enhance similarity searches in metric spaces. Despite these advancements, the issue of space overlap in index structures persists, leading to increased node accesses, longer search times, and higher computational costs—sometimes even exceeding those of sequential scans. Rather than resorting to solutions that eliminate all elements within the overlap area or minimize partitioning radii—both of which can lead to information loss—there is a critical need for a new mechanism to more effectively quantify, measure, and address the overlap problem. Such a mechanism should incorporate advanced metrics for accurately detecting overlaps and sophisticated methods for evaluating them during the partitioning process. To bridge this significant research gap, we propose three innovative heuristics designed to quantify and strategically manage data space partition overlap:

\begin{enumerate}
    \item Volume-Based Method (VBM): This heuristic provides a detailed measure of overlap by calculating the intersection volume between partitions. It offers a nuanced understanding of spatial relationships, enabling reliable quantification of overlap areas and optimizing partition boundaries. 
    \item Distance-Based Method (DBM): This approach evaluates overlap by analyzing the distance between partition centers and their radii. By focusing on the geometric relationships between partitions, this method ensures a computationally efficient process without sacrificing accuracy. 
    \item Object-Based Method (OBM): Notable for its practicality, this heuristic uses the count of objects shared between partitions as a straightforward yet insightful indicator of overlap. It facilitates a more intuitive interpretation of data space dynamics by directly linking overlap to data distribution.
 \end{enumerate}
 
This paper is organized as follows: Section \ref{sec2} offers an overview of the essential background information. Section \ref{sec3} reviews relevant existing works. Section \ref{sec4} presents the proposed heuristics in detail. Section \ref{sec5} analyzes the comparative experimental results. Finally, Section \ref{sec6} concludes the study and outlines future research directions.

\section{Theoretical Background}\label{sec2}

The metric space is a mathematical structure, introduced as a universal abstraction for data \cite{mao2015overcoming}, excels by relying solely on a distance function that satisfies the triangle inequality, thereby eliminating the need for specific content or intrinsic structure \cite{mao2016pivot, zezula2006similarity}. This versatility allows it to effectively index a wide variety of data types seamlessly.\\

\begin{definition}[Metric Space]

A metric space \(\mathscr{M}\) is defined as a pair \((\mathcal{S}, d)\), where \(\mathcal{S}\) denotes a domain of objects, and \(d: \mathcal{S} \times \mathcal{S} \rightarrow \mathbb{R^+}\) represents a distance function to quantify the “similarity” between objects within \(\mathcal{S}\). This distance function typically satisfies four distinct properties:

\begin{align*}
\forall (x, y) \in \mathcal{S},~ & d(x, y) \geq 0 & \text{(p$_1$: non-negativity)}, \\
\forall (x, y) \in \mathcal{S},~ & d(x, y) = d(y, x)   & \text{(p$_2$: symmetry)}, \\
\forall (x, y) \in \mathcal{S},~ & d(x, y) = 0 \Leftrightarrow x=y  &  \text{(p$_3$: identity)}, \\
\forall (x, y,z) \in \mathcal{S},~ &  d(x, y) + d(y, z) \geq d(x, z)  & \text{(p$_4$: triangle inequality)}. 
\end{align*}

\end{definition}

According to \cite{dohnal2004indexing}, the problem of indexing metric spaces can be formally stated as follows:\\

\begin{definition}[Problem of Data Indexing]
Let \( \mathscr{M} = (\mathcal{S}, d) \) be a metric space. Consider a subset \( \mathcal{X} \subseteq \mathcal{S} \), representing a set of objects, commonly referred to as the dataset, within which we perform searches. The problem of indexing metric spaces involves determining how to preprocess or structure the objects in \( \mathcal{X} \) so that similarity queries can be answered efficiently.
\end{definition}

\begin{figure}[h]
    \centering   
    \begin{subfigure}[b]{0.4\textwidth}
        \centering
        \includegraphics[width=\textwidth]{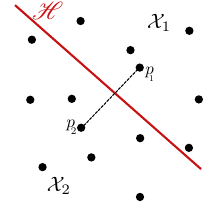}
        \caption{GH partitioning $\mathscr{H}(\mathcal{S}, d, p_1, p_2)$.}
        \label{fig:GH}
    \end{subfigure}
    \hfill
    \begin{subfigure}[b]{0.4\textwidth}
        \centering
        \includegraphics[width=\textwidth]{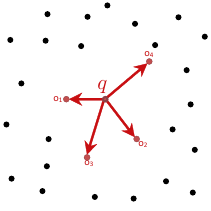}
        \caption{$k$NN query $k-\mathcal{NN} (\mathcal{S}, d, q, k)$.}
        \label{fig:Ball}
    \end{subfigure}
\end{figure}

As mentioned in Section \ref{sec1}, indexing techniques in this context typically involve partitioning a set \( \mathcal{X} \subseteq \mathcal{S} \) of objects in the metric space \( \mathscr{M} = (\mathcal{S}, d) \) into hierarchical regions centered around pivots using various partitioning techniques. One widely used method is Generalized Hyperplane (GH) partitioning, which is commonly employed in these indexes \cite{zezula2006similarity, chen2022indexing}. In GH partitioning, two specific objects \( (p_1, p_2) \in \mathcal{X} \) are chosen arbitrarily or specifically. The dataset \( \mathcal{X} \) is then divided into subsets \( \mathcal{X}_1 \) and \( \mathcal{X}_2 \) based on Definition (\ref{def:GH}) and as illustrated in Figure \ref{fig:GH}.\\

\begin{definition}[Generalized Hyperplane Partitioning]\label{def:GH}
Let \( \mathscr{M} = (\mathcal{S}, d) \) be a metric space, where \( \mathcal{X} \subseteq \mathcal{S} \) represents the set of objects to be indexed. Assume \( (p_1, p_2) \in \mathcal{S} \) are two pivot objects such that \( d(p_1, p_2) > 0 \). The generalized hyperplane

\begin{equation}
    \mathscr{H}(\mathcal{S}, d, p_1, p_2) = \{o \in \mathcal{S} : d(p_1, o) = d(p_2, o)\}
\end{equation}

partitions the space into two regions as follows:

\begin{align}
    \Biggl\{
    \begin{array}{ll}
    \mathcal{X}_1 &= \{ o \in \mathcal{X} : d(o, p_1) \leq d(o, p_2) \}\\
    &\\
    \mathcal{X}_2 &= \{ o \in \mathcal{X} : d(o, p_1) \geq d(o, p_2) \}
    \end{array}
\end{align}

\end{definition}

In metric space theory, a similarity query is defined by a query object \( q \) and a specified proximity criterion, typically expressed as a distance. The objective of such queries is to retrieve objects \(\{o_1, \dots, o_n ~|~ o_i \in \mathcal{X}\}\) that satisfy the proximity conditions relative to the query object \( q \) \cite{benrazek2021internet}. Among the various types of similarity queries, the $k$ nearest neighbor ($k$-NN) query is the most popular and widely used \cite{guzun2020high, moutafis2021algorithms}.

The $k$-NN query retrieves the \( k \) closest objects to a given query object \( q \) based on their distance in the metric space. If the dataset, contains fewer than \( k \) objects (i.e., $|\mathcal{X}| \leq k$), the query returns the entire dataset (i.e., $|A|=|\mathcal{X}|$) \cite{dohnal2004indexing}. This query type is formalized by the following definition.\\

\begin{definition}[$k$-Nearest Neighbor Query]
Let \( \mathscr{M} = (\mathcal{S}, d) \) a metric space, $q \in \mathcal{S}$ a query point, and $k \in \mathbb{N}$ the required number of answers.

\begin{equation}
\begin{split}
k-\mathcal{NN} (\mathcal{S}, d, q, k) =& \{A ~|~ A \subseteq \mathcal{X} \subseteq \mathcal{S}, |A| = k ~\wedge \\
 & \wedge~ \forall (o_i \in A)~\forall (o_j \in \{\mathcal{X} \setminus A\}) : d(q, o_i) \leq d(q, o_j)\}
\end{split}
\end{equation}

\end{definition}

\begin{figure}[h]
    \centering
    \includegraphics[width=0.65\linewidth]{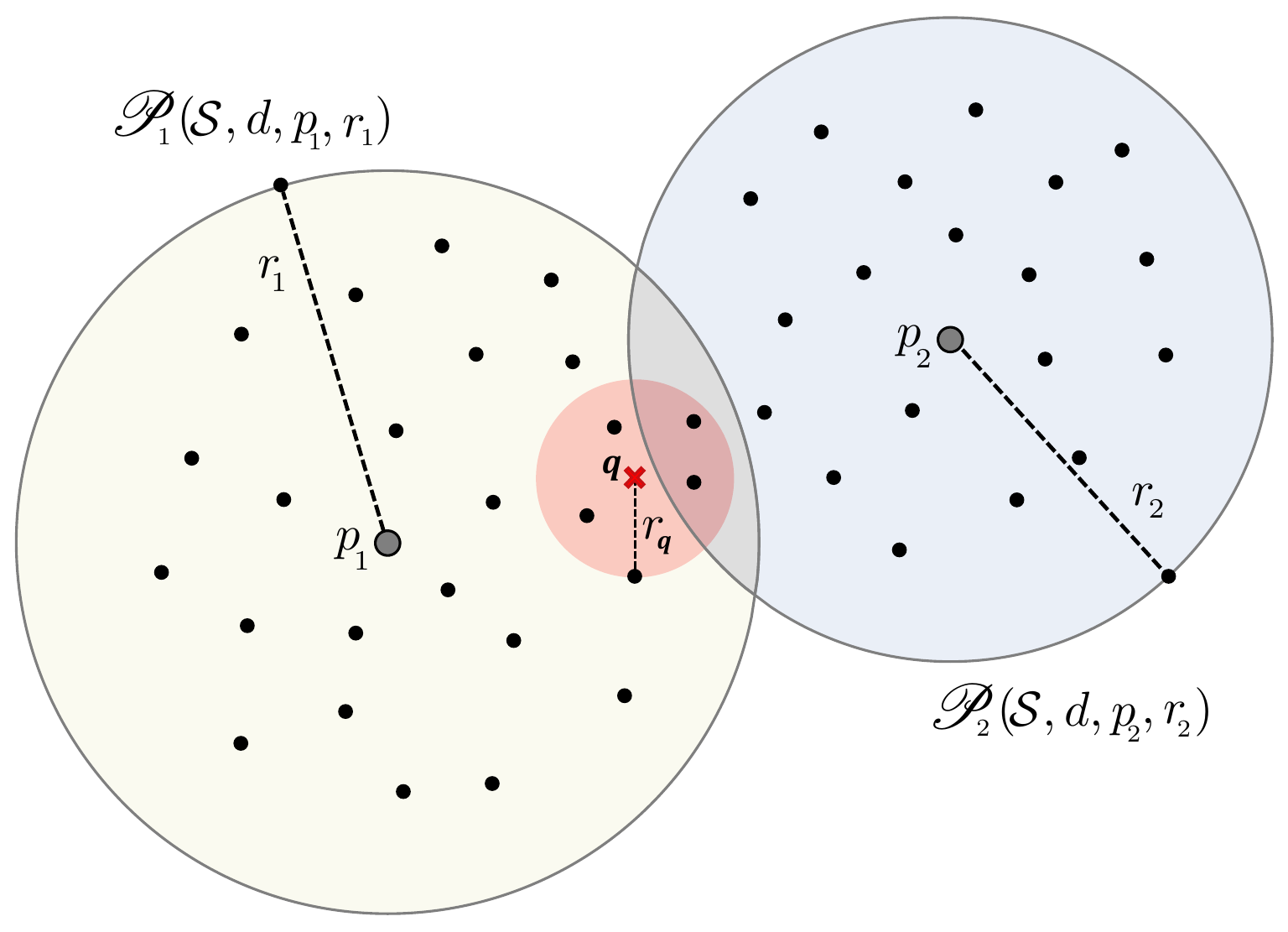}
    \caption{An example of overlapping partitions.}
    \label{fig:overlapping}
\end{figure}

During the partitioning process, distant points can sometimes be assigned to partitions, leading to a significant increase in their radii (\( r_i \)). Since these partitions are modeled as hyperballs (see Definition (\ref{def:hyperballs})), the inclusion of such outlier points can substantially expand the partition's coverage, potentially causing overlap in areas of the data space that do not reflect the actual data distribution. Theoretically, overlap between two partitions occurs when the farthest point within a partition (determined by its \( r_i \) value) lies farther from the partition's reference point (\( p_i \)) than the reference points are from each other. This scenario means that any query (\( q \in \mathcal{S} \)) within the overlapping space must be processed in each overlapping partition, even if some partitions contain no relevant data at the query location. As a result, distance calculations increase, search response times are prolonged, and computational costs rise, sometimes surpassing those of sequential scans, especially in cases of significant overlap. This ultimately impedes overall performance. For example, as illustrated in Figure \ref{fig:overlapping}, every partition intersected by the query sphere (\(\mathcal{P}_1\) and \(\mathcal{P}_2\)) must be searched, regardless of whether they contain relevant data points.\\

\begin{definition}[Hyperball]\label{def:hyperballs}
Let \( \mathscr{M}(\mathcal{S}, d) \) a metric space, \( \mathcal{X} \subseteq \mathcal{S} \) represents the set of objects. Consider \( p \in \mathcal{S} \) as a pivot object, and \( r \in \mathbb{R}^+ \) as the radius of coverage. The notation \( \mathscr{P}(\mathcal{S}, d, p, r) \) defines a closed ball that defined analogously the partitions, where:

\begin{equation}
    \mathscr{P}(\mathcal{S}, d, p, r) = \{o \in \mathcal{X} : d(o, p) \leq r \}
\end{equation}

\end{definition}

\section{Related Works}\label{sec3}

In the field of data indexing, numerous techniques have been developed to enhance query efficiency and retrieval speed. These challenges are particularly critical in modern data management systems, especially in dynamic and complex environments like the IoT. Various indexing techniques have been proposed to address efficient data retrieval in such settings, with several noteworthy structures standing out. One such technique involves Voronoi Diagram-based structures, such as GNAT \cite{brin1995near}, which leverage Voronoi Diagrams to address node overlaps through static partitioning. While this method effectively prevents node overlaps, it struggles in dynamic environments, particularly when maintaining non-overlapping partitions during real-time updates. The EGNAT \cite{navarro2011fully}, a dynamic variant of GNAT and a generalization of the GH-Tree \cite{uhlmann1991satisfying}, introduces flexible node organization to manage updates following an initial bulk load. However, it continues to face challenges in maintaining the non-overlapping principle of Voronoi diagrams \cite{moriyama2021vd}, and its search operations can be computationally intensive \cite{marin2007searching}.

In contrast to static structures, the M-Tree \cite{ciaccia1997m} and its variations, such as the M$^+$-Tree \cite{zhou2003m+} and BM$^+$-Tree \cite{zhou2005bm+}, are known for their balanced structures and effective handling of dynamic updates. Nevertheless, these structures encounter challenges with overlapping nodes at the same level, especially in higher-dimensional spaces, where computational overhead can increase significantly. The Slim-Tree \cite{traina2000slim} builds on the M-Tree by introducing innovative algorithms like the Slim-Down technique to minimize node overlap. However, the effectiveness of overlap reduction varies depending on the characteristics of the data distribution, and the need for object reinsertion during updates adds further complexity. OMNI technology \cite{traina2001similarity, traina2007omni}, a notable alternative to the M-Tree, significantly improves search efficiency by using strategically placed vantage points to increase the likelihood of node elimination. However, it still struggles with reducing the number of intersecting nodes, which can negatively impact overall search performance

Hybrid techniques, such as the IM-Tree \cite{kouahla2012new}, XM-Tree \cite{kouahla2019xm}, and NOBH-Tree \cite{pola2014nobh}, combine multiple strategies to mitigate overlap by utilizing pivot-based partitioning and non-overlapping regions. The IM-Tree and XM-Tree are hybrid memory-based structures designed to address overlap issues in spatial partitioning. The IM-Tree partitions space around two pivots into five non-overlapping regions, but encounters difficulties when distant regions accumulate too many objects. The XM-Tree addresses this by eliminating partitions without objects, introducing an external limiting area, and using linked lists for storing distant objects. The NOBH-Tree, another hybrid structure, combines Voronoi-shaped and ball-shaped partitions, utilizing two pivots, covering radii, and a hyperplane to create non-overlapping regions. It introduces a sixth region using the hyperplane to manage overlaps caused by the ball-shaped partitions, allowing for flexible tree compositions while ensuring that partitions meet the criteria for non-disjoint regions and symmetric space division relative to the pivots. Despite their potential, these hybrid structures involve complex partitioning and management strategies and do not fully resolve overlap issues, thus limiting their scalability and efficiency in large-scale deployments.

Recently, the BCCF-tree, introduced by \cite{benrazek2020efficient}, presented an innovative approach to real-time IoT data indexing, improving k-NN searches and achieving balanced partitioning using the k-means algorithm. Its extension, the ABCCF-tree, addresses unique labeling challenges in Internet of Video Things (IoVT) networks \cite{benrazek2023tree, allele2021automatic, benrazek2022introduction}. Another variant, the QCCF-tree \cite{Khettabi2021QCCF}, divides space into four regions and enhances query retrieval by enabling concurrent searches at each node, demonstrating superior performance in both construction and similarity searches. However, despite these advancements, these structures still struggle with overlap issues, particularly when managing large-scale datasets.

While the reviewed structures have shown effectiveness in various data environments, they continue to face significant challenges in efficiently managing node overlaps in large databases. Overlap complicates index construction and maintenance, and it undermines query performance by necessitating additional computations and data traversals. These ongoing challenges highlight a clear research gap: the need for a new mechanism that can more effectively quantify, measure, and address the overlap problem. Bridging this gap is essential for improving the efficiency of data indexing techniques.

\section{Proposed Solution}\label{sec4}

Our proposed solution divided into three successive stages. (i) Preprocessing, (ii) overlapping estimation, and (iii) decision-making and indexing.

\subsection{Preprocessing}

To enhance indexing structures, we propose applying a clustering algorithm to the data before indexing. The goal is to organize similar data into compact groups, facilitating efficient indexing and retrieval processes. In our approach, DBSCAN (Density-Based Spatial Clustering of Applications with Noise) \cite{ester1996density} is selected as the most suitable algorithm, supported by its proven effectiveness in previous studies \cite{kemouguette2021cost, khettabi2022clustering}. DBSCAN identifies clusters of arbitrary shapes based on density, dynamically adjusting without the need to pre-determine the number of clusters, unlike $k$-means. This method relies on two key parameters: (i) \( \epsilon \), the radius of the spherical neighborhood around each object \( o \in \mathcal{X} \) (see Definition (\ref{def:epsilon})), and (ii) \texttt{MinPts}, the minimum number of objects required for a neighborhood to be considered a valid cluster.\\

\begin{definition}[$\epsilon$-Neighborhood of an Object]\label{def:epsilon}
Let \( \mathscr{M} = (\mathcal{S}, d) \) be a metric space, and let \( \mathcal{X} \subseteq \mathcal{S} \) represent the set of objects to be grouped. The $\epsilon$-neighborhood of an object $o$, denoted as $\mathcal{N}_{\epsilon}(o)$, is defined as:    

\begin{equation}
    \mathcal{N}_{\epsilon }(o)=\{q \in \mathcal{X} ~|~ d(o, q) \leq \epsilon\}
\end{equation}
\end{definition}

Despite DBSCAN's ability to cluster without requiring centroids or radii for each group, it does not fully align with all our specific needs. Therefore, we will enhance the DBSCAN algorithm (Algorithm \ref{alg:dbscan})\footnote{Please note that the function “ExpandCluster” from Algorithm \ref{alg:dbscan} remains unchanged. We used it as is and therefore did not include its details here. For further details on “ExpandCluster”, refer to \cite{ester1996density}.} to better suit our unique requirements in the subsequent stages of this research.

\begin{algorithm}[H]
\caption{DBSCAN Clustering Algorithm}
\label{alg:dbscan}
\begin{algorithmic}[1]
\Require $\left(
                  \begin{array}{l}
                     \mathcal{X} \subseteq \mathcal{S}, \\
                     d : \mathcal{S} \times \mathcal{S} \to \mathbb{R}^+,\\
                     \epsilon \in  \mathbb{R}^+, \\
                     \texttt{MinPts} \in  \mathbb{N}
                  \end{array}
               \right)$
\Ensure $\left(
                  \begin{array}{l}
                     \mathcal{\{P\}} \in \mathcal{X}, \\
                     \{p\} \in \mathcal{S}, \\
                     \{r\} \in \mathbb{R}^+
                  \end{array}
               \right)$
\State \text{ClusterId} $\gets$ \text{nextId(NOISE)}
\For{all $o \in \mathcal{X}$}
    \If{$o$.ClId = UNDEFINED}
        \If{ExpandCluster($\mathcal{X}$, $o$, ClusterId, $\epsilon$, \texttt{MinPts})}
            \State ClusterId $\gets$ nextId(ClusterId)
        \EndIf
    \EndIf
\EndFor
\State $\{\mathcal{P}_i\}^{ClusterId}_{i=1} = \{\forall o \in \mathcal{X} ~|~ o\text{.ClId} = i\}$ \Comment{The set of partitionings.}
\State $\{p_i\}_{i=1}^{ClusterId} = \left\{ \frac{1}{|\mathcal{P}_i|} \sum_{o \in \mathcal{P}_i} o \right\}$ \Comment{The set of pivots of partitionings.}
\State $\{r_i\}_{i=1}^{ClusterId} = \max_{o \in \mathcal{P}_i}\{d(p_i, o)\}$ \Comment{the set of radii of partitionings.}

\end{algorithmic}
\end{algorithm}

\subsection{Overlapping Estimation}

In this subsection, we introduce our innovative heuristics designed to minimize the impact of the overlap space problem discussed in Section \ref{sec2}. These heuristics aim to enhance metric space indexing without compromising query retrieval accuracy or the overall quality of the index.

\subsubsection{Volume-Based Method (VBM)}

\begin{figure}[H]
    \centering   
    \begin{subfigure}[b]{0.49\textwidth}
        \centering
        \includegraphics[width=\textwidth]{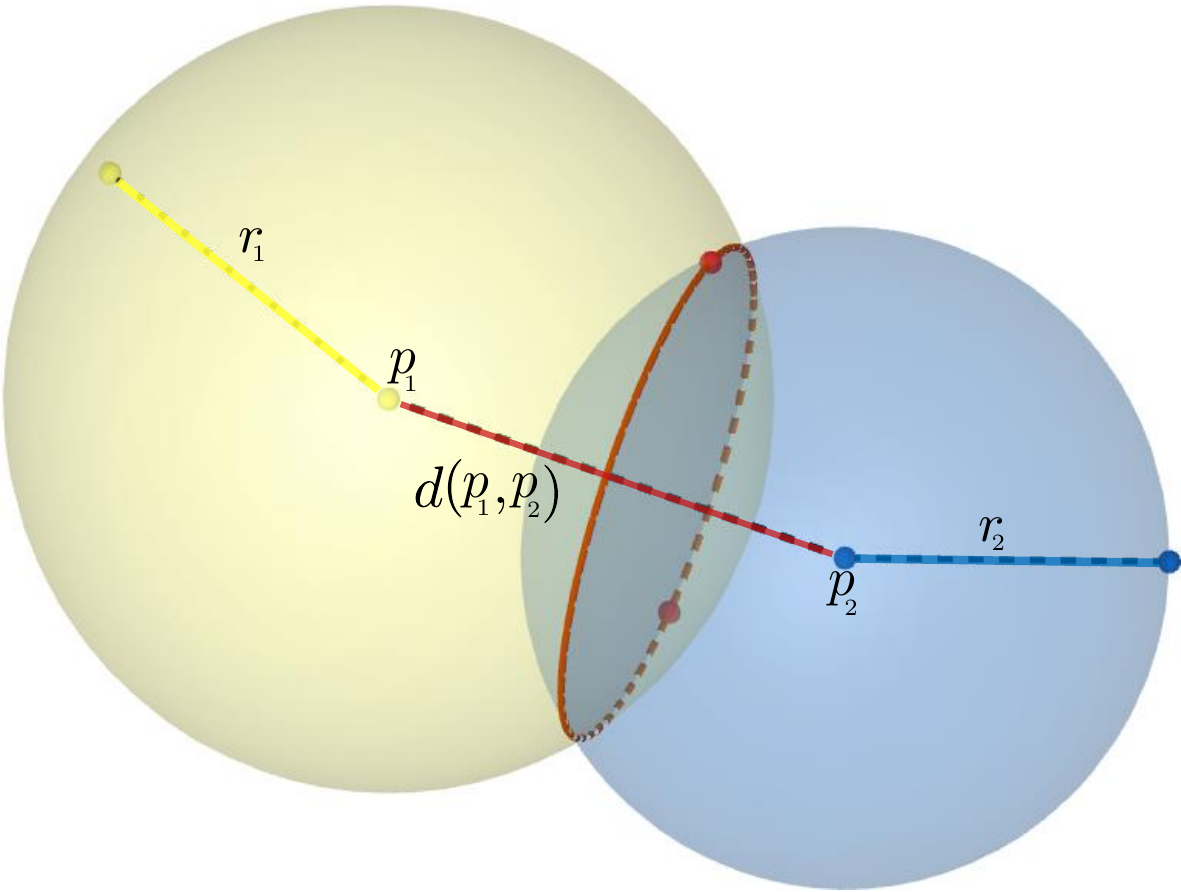}
        \caption{Projection 2D.}
        \label{fig:2D}
    \end{subfigure}
    \hfill
    \begin{subfigure}[b]{0.49\textwidth}
        \centering
        \includegraphics[width=\textwidth]{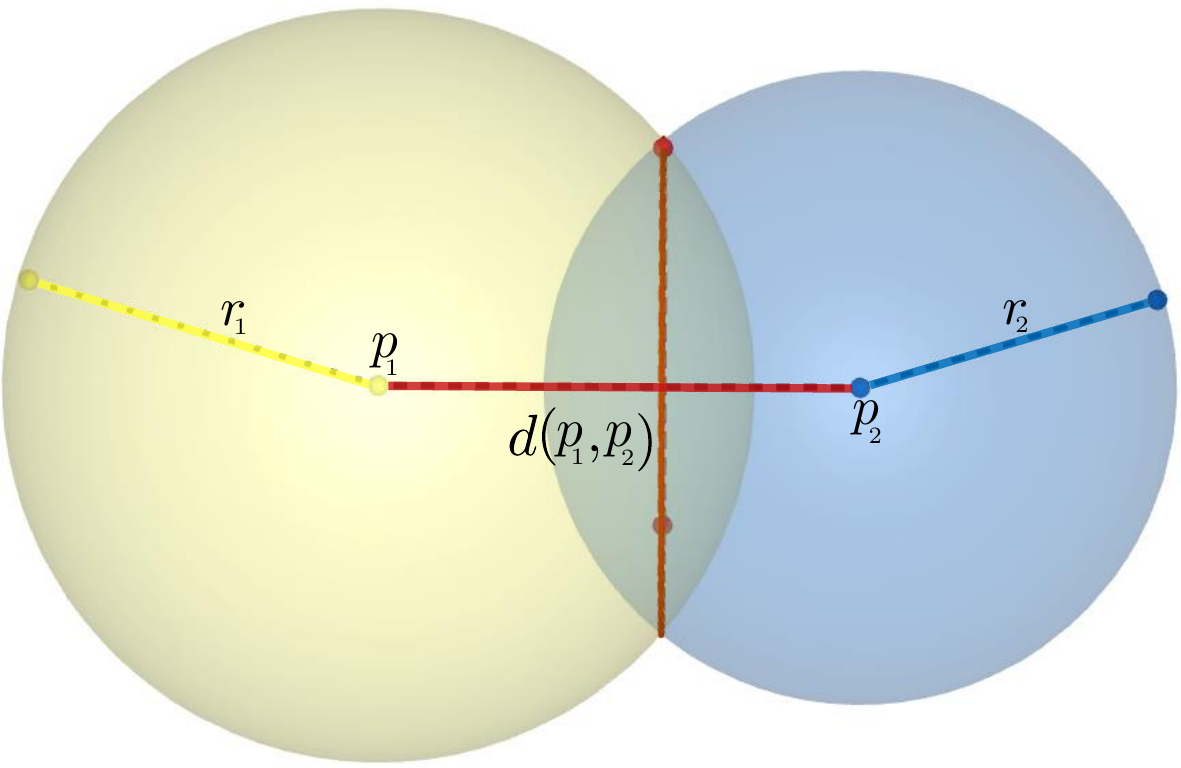}
        \caption{Projection 3D.}
        \label{fig:3D}
    \end{subfigure}
    \caption{Geometry of the intersection of two hyperballs.}
    \label{fig:intersection}
\end{figure}

The first proposed heuristic is the Volume-Based Method (VBM). This approach provides a detailed measure of overlap by calculating the intersection volume between partitions. As outlined in Section \ref{sec2}, partitions in a metric space are modelled as hyperballs (see Definition (\ref{def:hyperballs})). To accurately assess the degree of partition overlap, we measure the overlapping volume between these corresponding hyperballs. The intersection volume between two hyperballs (see Figure \ref{fig:intersection}) depends solely on the distance between their centers and the sizes of their radii, as outlined in Definition (\ref{def:intersection}).

\begin{definition}\label{def:intersection}

    Let \( \mathscr{M}(\mathcal{S}, d) \) a metric space. Consider two partitions \(\mathscr{P}_1(\mathcal{S}, d, p_1, r_1)\) and \(\mathscr{P}_2(\mathcal{S}, d, p_2, r_2)\), representing hyperballs centered at \(p_1\) and \(p_2\) with radii \(r_1\) and \(r_2\), respectively. The  overlap volume (\(V\)) and overlap volume rate (\(V_{\cap}\)) are defined as:

\begin{equation}
    V =
\begin{cases}
0 & \text{if } d(p_1, p_2) \geq r_1 + r_2 \\
\min\{V_{\mathscr{P}_1}, V_{\mathscr{P}_2}\} & \text{if } d(p_1, p_2) \leq |r_1 - r_2| \\
\sum_{i=1}^{2} {V_{\bigtriangledown}(r_i,p_i)} & \text{otherwise}
\end{cases}
\end{equation}

\begin{equation}
    V_{\cap} =
\begin{cases}
0 & \text{if } d(p_1, p_2) \geq r_1 + r_2 \\
1 & \text{if } d(p_1, p_2) \leq |r_1 - r_2| \\
\frac{\sum_{i=1}^{2} {V_{\bigtriangledown}(r_i,p_i)}}{V_{\mathscr{P}_1} + V_{\mathscr{P}_2}}& \text{otherwise}
\end{cases}
\end{equation}

Where:
\begin{itemize}
    \item \(V_{\mathscr{P}_1}\) and \(V_{\mathscr{P}_2}\) are the volumes of hyperballs \(\mathscr{P}_1\) and \(\mathscr{P}_2\), respectively, as defined in Definition (\ref{def:Volume_Hyperball}).
    \item \(V_{\bigtriangledown}(r_i,c_i)\) is the volume of the spherical cap of the hyperball \(\mathscr{P}_i\) and is calculated using Definition (\ref{def:Hyperspherical_Cap}).
\end{itemize}

\end{definition}

\begin{definition}[Volume of a Hyperball] \label{def:Volume_Hyperball}
Let \( \mathscr{M}(\mathcal{S}, d) \) be a metric space. Consider a partition \(\mathscr{P}_i(\mathcal{S}, d, p_i, r_i)\), representing an $n$-dimensional hyperball centered at \(c_i\) with radius \(r_i\). The volume \(V_{\mathscr{P}_i}\) is given by \cite{li2010concise}:

 \begin{equation}
     V_{\mathscr{P}_i} = \frac{\pi^{n/2}}{\Gamma(\frac{n}{2} + 1)} \cdot r_i^n
 \end{equation} 

where $\Gamma$ denotes the gamma function \cite{artin2015gamma}:

\begin{equation}
    \Gamma(x) = \int_{0}^{\infty} t^{x-1} e^{-t} \, dt
\end{equation}
    
\end{definition}

\begin{figure}[h]
    \centering
    \includegraphics[width=0.6\linewidth]{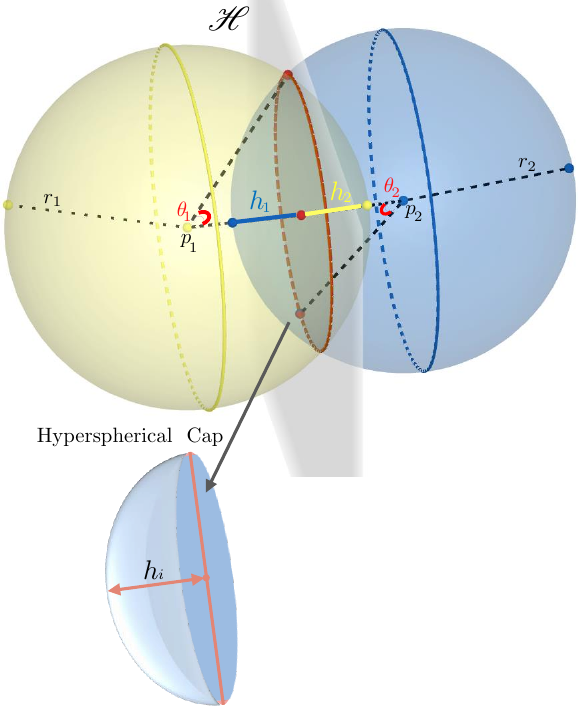}
    \caption{An example of a spherical cap.}
    \label{fig:Caps}
\end{figure}

As demonstrated in Definition (\ref{def:intersection}), the volume of the intersection between two hyperballs corresponds to the combined volume of the spherical caps formed by a hyperplane cutting through them (see Figure \ref{fig:Caps}).

To determine the \( n \)-dimensional volume \( V_{\bigtriangledown}(r_i, p_i) \) of a hyperspherical cap with height \( h_i \) and radius \( r_i \), we apply the methodology described in \cite{li2010concise}. The formula for calculating this volume is formalized by the following definition:\\

\begin{definition}[Volume of Hyperspherical Cap] \label{def:Hyperspherical_Cap}
Let \( \mathscr{M}(\mathcal{S}, d) \) denote a metric space. Consider two partitions, \( \mathscr{P}_i(\mathcal{S}, d, p_i, r_i) \) and \( \mathscr{P}_j(\mathcal{S}, d, p_j, r_j) \), representing \( n \)-dimensional hyperballs centered at $p_i$ and $p_j$ with radius $r_i$ and $r_j$, respectively. The volume of a hyperspherical cap of \( \mathscr{P}_i \) with height \( h_i \) is calculated using the formula from \cite{li2010concise} as follows:

\begin{equation}
    V_{\bigtriangledown}(r_i,p_i) = \frac{\pi^{\frac{n-1}{2}} r{_i}^n}{\Gamma(\frac{n+1}{2})} \int_{0}^{\arccos{(\frac{r{_i}-h{_i}}{r{_i}})}} \sin^n{(\theta{_i})}d\theta_i
\end{equation}

Where:
\begin{itemize}
    \item $h_i$ is the height of the cap, calculated by:
    
    \begin{equation}
        h_i = r_i(1-cos(\theta_i))
    \end{equation}

    \item \( \theta_i \) is the polar angle between the rays from the center \( p_i \) of the sphere to the apex of the cap (the pole) and the edge of the disk forming the base of the cap, calculated as:

    \begin{equation}
        \theta_i = \cos^{-1} \frac{r{_i}{^2}+d(p_i, p_j)^2-r{_j}{^2}}{2r_id(p_i, p_j)}
    \end{equation}
    
\end{itemize}
\end{definition}

\subsubsection{Distance-Based Method (DBM)}

The Distance-Based Method (DBM) is a simpler heuristic for evaluating overlap between partitions by calculating the distance between their centers and radii. Unlike the VBM, the DBM focuses exclusively on distance relationships, ensuring a computationally streamlined process without compromising accuracy. As illustrated in Figure \ref{fig:Caps}, the overlap distance between partitions is represented by the combined height of the caps. Calculating the rate of this overlap, as defined in Definition (\ref{def:intersection}), is crucial for accurately quantifying the degree of overlap. This quantification helps optimize partition boundaries and enhances the efficiency of the indexing structure.\\

\begin{definition}[Overlapping Distance Rate]\label{def:Distanceintersection}

    Let \( \mathscr{M}(\mathcal{S}, d) \) a metric space. Consider two partitions, \(\mathscr{P}_1(\mathcal{S}, d, p_1, r_1)\) and \(\mathscr{P}_2(\mathcal{S}, d, p_2, r_2)\), representing hyperballs centered at \(c_1\) and \(c_2\) with radii \(r_1\) and \(r_2\), respectively. The distance rate of their intersection ($D_{\cap}$) is calculated as follows:

\begin{equation}
    D_{\cap} =
\begin{cases}
0 & \text{if } d(p_1, p_2) \geq r_1 + r_2 \\
1 & \text{if } d(p_1, p_2) \leq |r_1 - r_2| \\
\frac{\sum_{i=1}^{2} {h_i}}{d(p_1, p_2)} & \text{otherwise}
\end{cases}
\end{equation}

Where $h_i$ represents the height of the cap of hyperball \(\mathscr{P}_i\) as defined in Definition (\ref{def:Hyperspherical_Cap}).

\end{definition}

\subsubsection{Object-Based Method (OBM)}

The Object-Based Method (OBM) is particularly practical, using the count of objects shared between partitions as a straightforward yet insightful indicator of overlap. This method allows for a more intuitive interpretation of data space dynamics by directly linking overlap to the actual distribution of data. Unlike previous methods that calculate overlapping space—potentially leading to inaccuracies in cases of large but empty overlaps—OBM addresses this issue by focusing on the number of overlapping objects, providing a more accurate representation of data overlap. The mathematical foundation of this approach is illustrated in Definition (\ref{def:OBM}).\\

\begin{definition}[Overlapping Objects Rate]\label{def:OBM}

Let \( \mathscr{M} = (\mathcal{S}, d) \) be a metric space, and let \( \mathcal{X} \subseteq \mathcal{S} \) represent the set of partitioned objects. Consider two partitions, \(\mathscr{P}_1(\mathcal{S}, d, p_1, r_1)\) and \(\mathscr{P}_2(\mathcal{S}, d, p_2, r_2)\), representing hyperballs centered at \(p_1\) and \(p_2\) with radii \(r_1\) and \(r_2\), respectively. The set \( A \) of objects located in the overlapping area of the two hyperballs is defined by:

\[A = \mathscr{P}_1(\mathcal{S}, d, p_1, r_1) \cap \mathscr{P}_2(\mathcal{S}, d, p_2, r_2)\]

Explicitly, this is defined as:
\[A = \{ o \in \mathcal{X} \mid d(o, p_1) \leq r_1 \text{ and } d(o, p_2) \leq r_2 \}\]

The rate \( A_{\cap} \) of objects in the overlapping area \( A \) is calculated as:

\[
A_{\cap} =
\begin{cases}
0 & \text{if } d(p_1, p_2) \geq r_1 + r_2 \\
1 & \text{if } d(p_1, p_2) \leq |r_1 - r_2| \\
\frac{|A|}{|\mathscr{P}_1|+|\mathscr{P}_2|} & \text{otherwise}
\end{cases}
\]

Here,
\begin{itemize}
    \item \( |A| \) denotes the number of objects in set \( A \).
    \item \( |\mathscr{P}_1| \) and \( |\mathscr{P}_2| \) represent the number of objects in partitions \( \mathscr{P}_1 \) and \( \mathscr{P}_2 \), respectively.
\end{itemize}
\end{definition}

\subsection{Decision-Making and Indexing}

After estimating the overlap between partitions, the next step is to create the indexes. We establish two thresholds, \( \xi_{min} \) and \( \xi_{max} \), to quantify the degree of overlap. Based on these thresholds, we categorize overlap into three levels: (1) Low overlap [0, \( \xi_{min} \)], (2) Medium overlap [\( \xi_{min} \), \( \xi_{max} \)], and (3) High overlap [\( \xi_{max} \), 1]. These classifications enable us to balance the trade-offs between index complexity and retrieval efficiency by applying tailored strategies to different levels of overlap, thereby ensuring optimal performance. By defining these thresholds, we can adapt our indexing approach to the specific characteristics of the data and the nature of the overlap.

\begin{itemize}
    \item Low overlap [0, \( \xi_{min} \)]: In cases of low overlap, we extract all objects from the partition with the smaller cap and merge them into the other partition. This results in two separate indexes, one for each partition. This minimal overlap can be efficiently managed by redistributing objects, which maintains clear separation between partitions and reduces redundant searches.
    
    \item Medium overlap [\( \xi_{min} \), \( \xi_{max} \)]: For medium overlap, we create three indexes: one for each partition, excluding the overlapping objects, and a third specifically for the overlapping objects. This third index is labeled as a neighbor of the original indexes (partitions) to ensure easy access during the search, particularly when the required number of objects has not yet been reached. This approach ensures a balance between merging and maintaining separate partitions. By isolating overlapping objects into a separate index, searches within each primary partition remain efficient while the overlap is accounted for in a dedicated index.
    
    \item High overlap [\( \xi_{max} \), 1]: In cases of high overlap, we merge the two partitions to create a single index structure for all objects. Significant overlap would lead to excessive redundancy and inefficient searches if the partitions were kept separate. Merging them into a single index simplifies the structure and optimizes retrieval by eliminating the need for extensive overlap management.
\end{itemize}

The thresholds \( \xi_{min} \) and \( \xi_{max} \) are crucial for classifying the degree of overlap. Their values directly influence how overlap is quantified and subsequently managed. Setting these thresholds requires careful consideration of the dataset's characteristics and the desired balance between indexing complexity and search efficiency. By fine-tuning these thresholds, we can develop an optimal indexing strategy that minimizes computational overhead while ensuring accurate and efficient data retrieval.

For our index structures, we employ the BCCF-tree \cite{benrazek2020efficient}, an efficient metric index structure designed for big IoT data retrieval. The BCCF-tree partitions the data space into disjoint regions using a recursive $k$-means algorithm, which is effective in minimizing overlap between partitions. However, the construction of the BCCF-tree can be complex due to the recursive nature of the $k$-means algorithm. By addressing this challenge through the steps outlined earlier, the intricate nature of the $k$-means algorithm becomes less critical, allowing for the use of a simpler partitioning technique. Therefore, we propose a refined approach utilizing GH partitioning, as defined in Definition \ref{def:GH}. GH partitioning is widely recognized for its effectiveness and relatively lower computational complexity compared to other methods. The complexity of hyperplane partitioning typically ranges from \(O(n \log(n))\) to \(O(n^2)\), depending on the specific implementation and the dimensionality of the data, where \(n\) represents the number of data points \citep{zezula2006similarity}.\\

\begin{definition}[BCCF-tree]\label{def:BCCF-tree}
    Let \( \mathscr{M}(\mathcal{S}, d) \) a metric space, and let \( \mathcal{X} \subseteq \mathcal{S} \) represent the set of indexed objects. The BCCF-tree structure is composed of two levels:

    \begin{itemize}
        \item Internal nodes $\mathcal{N}$ is a sextuple:

        \begin{equation}
            (p_1, p_2, r_1, r_2, \mathcal{N}_L, \mathcal{N}_R)
      \in \mathcal{X} \times \mathcal{X} \times \mathbb{R}^+ \times {\mathbb{R}^+} \times  \mathcal{N} \times \mathcal{N}
        \end{equation}
   where:
   \begin{itemize}
      \item $(p_1, p_2)$ are two pivots, with $d(p_1, p_2) > 0$. 
      \item $(r_1, r_2)$ represent the distances to the farthest object in the sub-tree rooted at that node $\mathcal{N}$ with respect to $p_1$ and $p_2$, respectively, i.e., $r_i = \max\{d(p_i, o), \forall o \in \mathcal{N}\}$.
      
      \item $(\mathcal{N}_L, \mathcal{N}_R)$ are two sub-indexes (left and right child nodes).
  \end{itemize}
  
  \item Leaf nodes, denoted as $E_C$, consist of a subset of the objects stored within a bucket.

   \begin{equation}
       E_C \subseteq \mathcal{X}
   \end{equation}

   where, $|E_C| \leq c_{\max}$, $c_{max}  = \sqrt{n}$, and $n = |\mathcal{X}|$
  
    \end{itemize}
\end{definition}

In this paper, we employ the BCCF-tree structure without modification, using the same construction algorithm detailed in \cite{benrazek2020efficient, benrazek2021internet}. However, during the search phase, we introduce a new algorithm to manage the multiple indexes generated. This algorithm selects the appropriate index or indexes based on their proximity to the query object (see Algorithm \ref{algo:close_indexes}). Once the closest indexes are identified, the $k$NN search algorithm from \cite{benrazek2020efficient, benrazek2021internet} is applied in parallel across all selected indexes, ensuring efficient and accurate query processing while leveraging parallelism for enhanced performance.

\begin{algorithm}[H]
\caption{Recherche $k$NN}
\label{algo:close_indexes}
\begin{algorithmic}[1]
   
    \Require        $\left(\begin{array}{l}
                     indexes \in \mathcal{N}, \\
                     q \in \mathcal{S}, \\
                     d : \mathcal{S} \times \mathcal{S} \to \mathbb{R}^+        \end{array}\right)$ \Comment{A set of indexes \& query object.}

    \Ensure $close\_indexes \in \mathcal{N}$ \Comment{Set of indexes closest to $q$.}
    \State \texttt{\textbf{/** STEP 01:} Estimation of the closest indexes to the query $q$ **/}
    \State $min\_index \gets 0$
    \State $min\_dist \gets \infty$
    \State $close\_index \gets None$
    \State $A \gets \varnothing$
    \For{$each~index \in indexes$)}
        \State $dist \gets d(index.center, q)$
        \If{$dist < min\_dist $}
            \State $close\_index \gets index$
            \State $min\_dist \gets dist$
        \EndIf
    \EndFor
    \State $close\_indexes.append(close\_index)$
    \For{$index \in close\_index.neighbors$} 
        \State $close\_indexes$.append($index$)
    \EndFor     
    \State \texttt{\textbf{/** STEP 02:} Run $k$NN search on the selected indexes **/}
    \For{\textbf{all} $index \in close\_indexes$} \textbf{in PARALLEL}
    \State $r_q \gets $ Estimated-$r_q$($index, q, d, r_q=\infty$) \Comment{Run the radius estimate $r_q$ of \cite{benrazek2021internet}}
    \State $A \gets k$NN-BCCF-index($index, q, k, d, r_q, A$) \Comment{Run $k$NN algorithm of \cite{benrazek2021internet}}
    \State $results$.append($A$)
    \EndFor 

    \State \texttt{\textbf{/** STEP 03:} Gather results from all parallel processes **/}
    \State $final\_results \gets Gather(results)$

    \State \Return $final\_results$
\end{algorithmic}
\end{algorithm}

\section{Simulation and Results}\label{sec5}

\subsection{Simulation Setup}
The prototype was implemented in Python on a workstation equipped with an Intel® Core™ i5-4200U CPU at 1.6 GHz, 4 GB of RAM, and running the Linux (Ubuntu) operating system.

Two datasets were used in the experiments. Table \ref{tab:datasets} provides an overview of the dataset characteristics, along with the parameters used in the DBSCAN algorithm ($\epsilon$ and \texttt{MinPts}) for each dataset. Additionally, the table includes the fixed overlap thresholds (\(\xi_{min}\) and \(\xi_{max}\)) used in the experiments.

\begin{itemize}
    \item \textbf{DB1}: This dataset consists of feature vectors from moving objects, obtained from an object tracking simulator in an IoVT environment using wireless cameras \cite{benrazek2023iovt}.
    \item \textbf{DB2}: Known as the Wearable Action Recognition Database (WARD)\footnote{WARD version 1.0: [\href{https://people.eecs.berkeley.edu/~yang/software/WAR/WARD1.zip}{download}]}, this dataset is a benchmark database used for recognizing human actions using a wearable motion sensor network \cite{yang2009distributed}.
\end{itemize}

\begin{table}[h!]
    \centering
    \begin{tabular}{|l|c|c|c|c|c|c|c|}
        \hline
        \textbf{Dataset} & \textbf{Size} & \textbf{Dimensions} & \textbf{$C_{max}$} &$\epsilon$& \texttt{MinPts}& $\xi_{min}$& $\xi_{max}$ \\ \hline
        Tracking         & 62,702                     & 20                  & 250 &248 & 250& \multicolumn{1}{c|}{\multirow{2}{*}{0.4}} &  \multicolumn{1}{c|}{\multirow{2}{*}{0.8}}               \\ 
        WARD             & 1,000,000                  & 5                   & 316 & 91 & 23 & &                 \\ \hline
    \end{tabular}
    \caption{Parameter values.}
    \label{tab:datasets}
\end{table}

To comprehensively evaluate the effectiveness of the proposed solution, a series of experiments were conducted, benchmarking its performance against the baseline BCCF-tree structure. The experiments were designed to assess the solution across three key dimensions:
(i) Structure evaluation, (ii) Construction cost, and (iii) Search efficiency.

\subsection{Structure Evaluation}

This section provides a detailed analysis of the quality of the indexes generated by the proposed heuristics—DBM, OBM, and VBM—applied to the Tracking and WARD datasets. The analysis primarily evaluates these heuristics based on their effectiveness in managing data space and optimizing tree structures. Key metrics assessed include the number of leaf nodes (or buckets) and the distribution of data within these nodes, the number and distribution of internal nodes at each index level, and the overall height of the tree.

\subsubsection{Tracking dataset}
\begin{figure}[h]
    \centering
    \begin{subfigure}{0.32\textwidth}
        \centering
        \includegraphics[width=\linewidth]{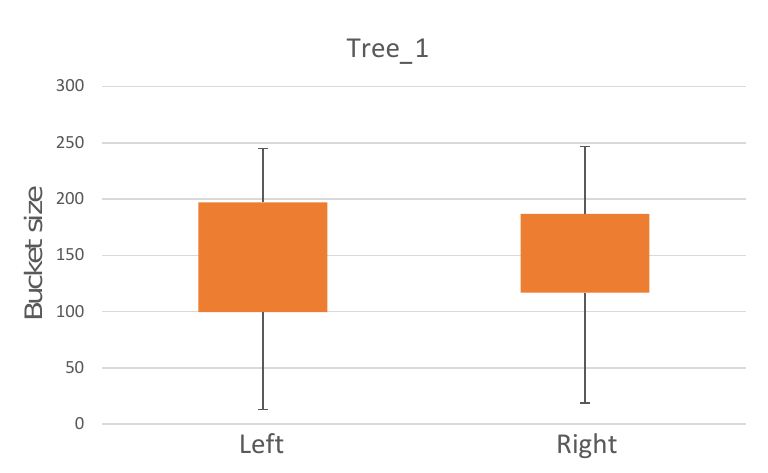}
    \end{subfigure}
    \hfill
    \begin{subfigure}{0.32\textwidth}
        \centering
        \includegraphics[width=\linewidth]{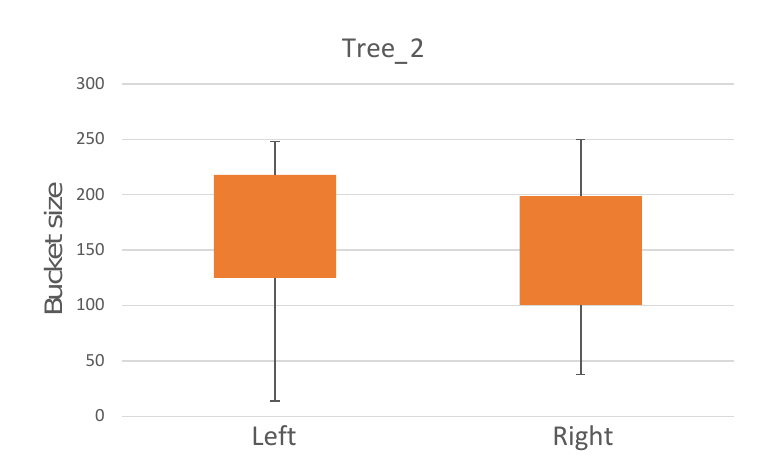}
    \end{subfigure}
    \hfill
    \begin{subfigure}{0.32\textwidth}
        \centering
        \includegraphics[width=\linewidth]{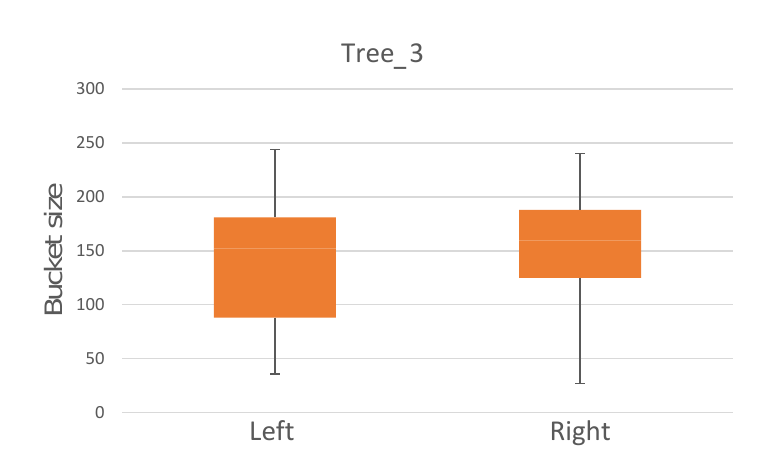}
    \end{subfigure}
    \vspace{1em} 
    \begin{subfigure}{0.33\textwidth}
        \centering
        \includegraphics[width=\linewidth]{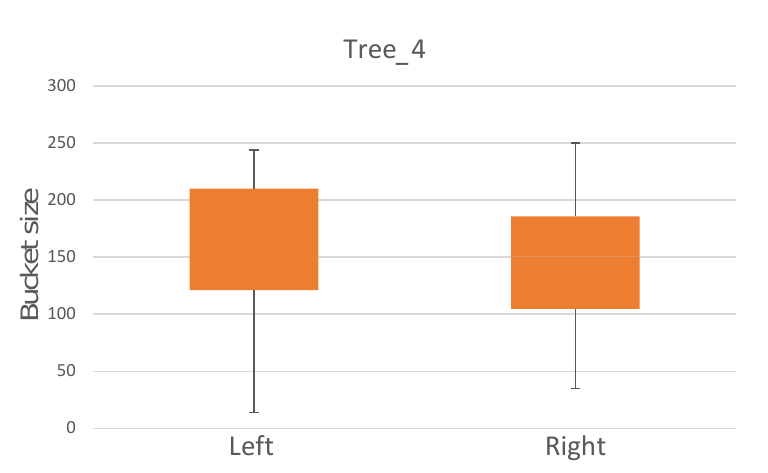}
    \end{subfigure}
    \begin{subfigure}{0.33\textwidth}
        \centering
        \includegraphics[width=\linewidth]{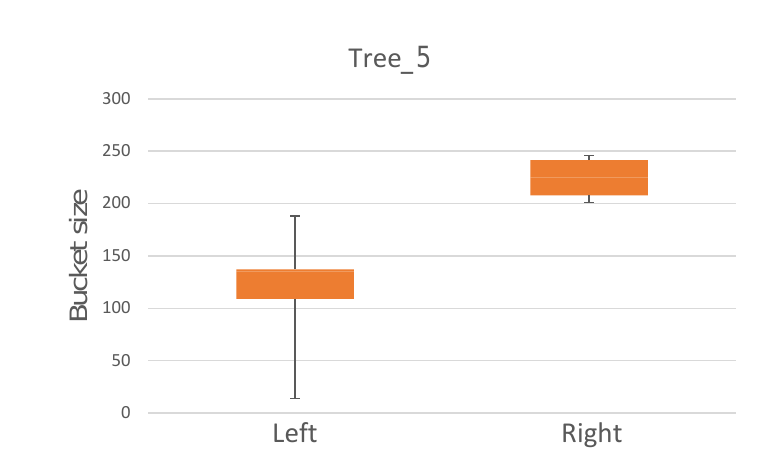}
    \end{subfigure}
    \caption{Bucket data distribution using DBM for Tracking datasets.}
    \label{fig:data_distribution_D_T}
\end{figure}

\begin{figure}[h]
    \centering
    \begin{subfigure}{0.32\textwidth}
        \centering
        \includegraphics[width=\linewidth]{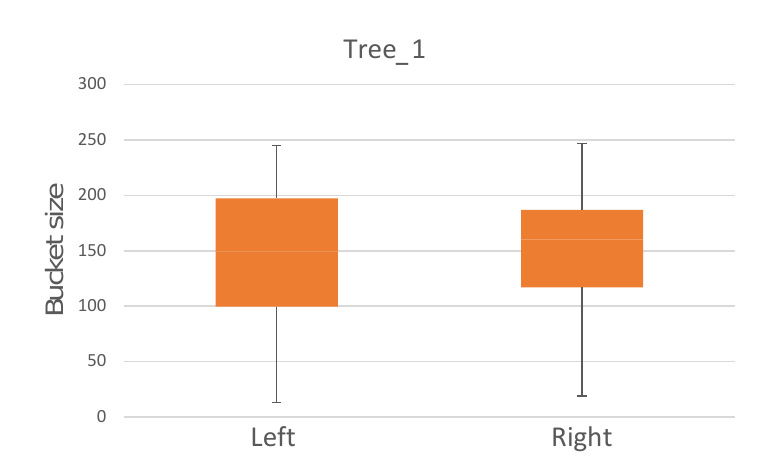}
    \end{subfigure}
    \hfill
    \begin{subfigure}{0.32\textwidth}
        \centering
        \includegraphics[width=\linewidth]{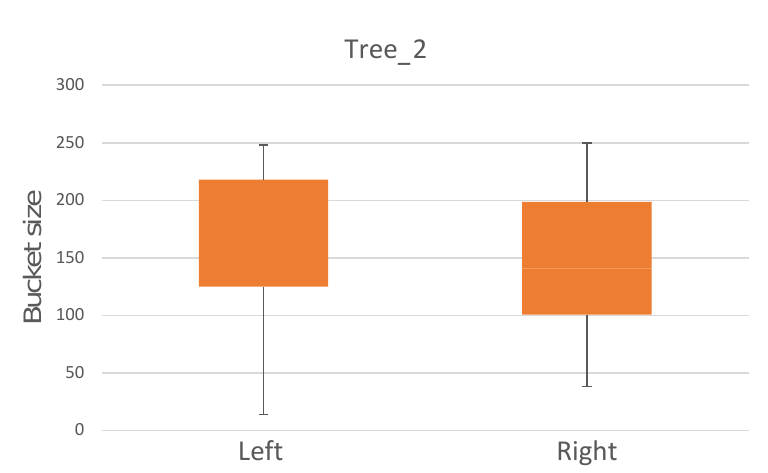}
    \end{subfigure}
    \hfill
    \begin{subfigure}{0.32\textwidth}
        \centering
        \includegraphics[width=\linewidth]{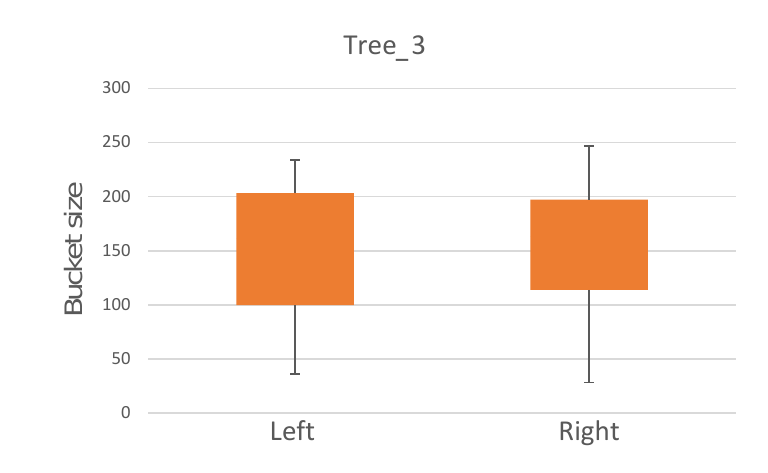}
    \end{subfigure}
    \vspace{1em} 
    \begin{subfigure}{0.33\textwidth}
        \centering
        \includegraphics[width=\linewidth]{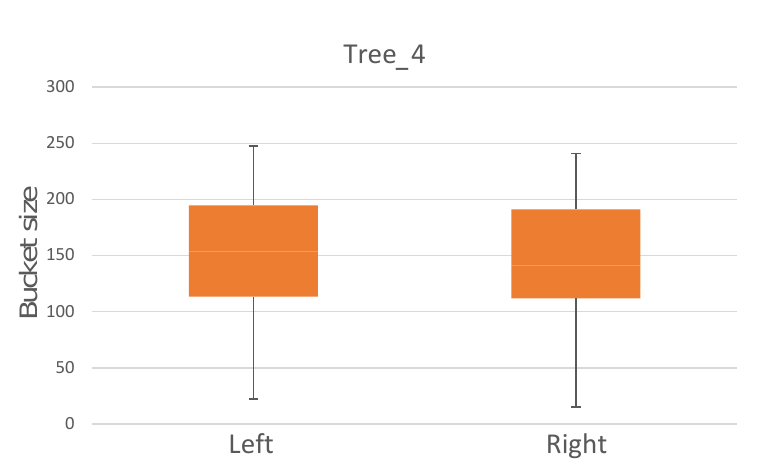}
    \end{subfigure}
    \hspace{3em}
    \begin{subfigure}{0.33\textwidth}
        \centering
        \includegraphics[width=\linewidth]{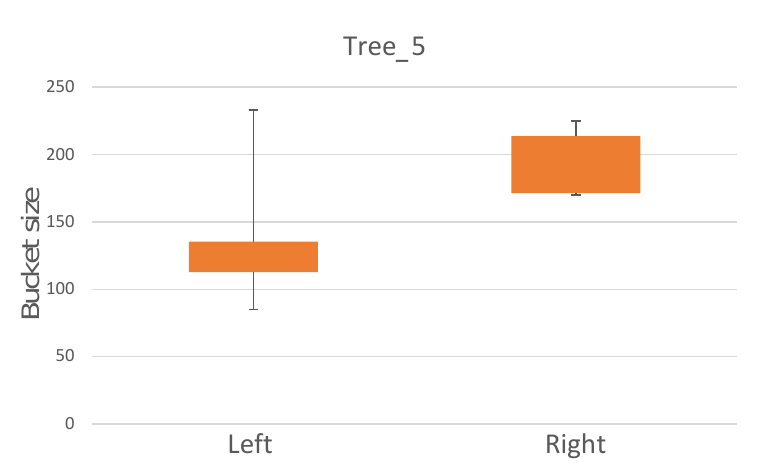}
    \end{subfigure}
    \caption{Bucket data distribution using OBM for Tracking datasets.}
    \label{fig:data_distribution_O_T}
\end{figure}

\begin{figure}[h]
    \centering
    \begin{subfigure}{0.32\textwidth}
        \centering
        \includegraphics[width=\linewidth]{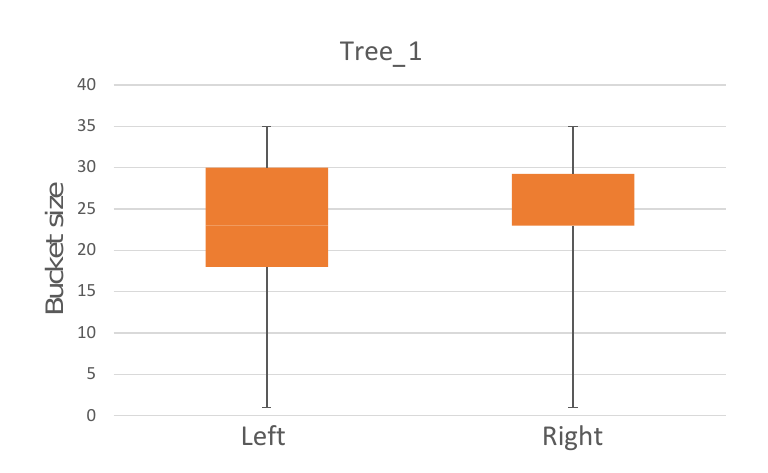}
    \end{subfigure}
    \hfill
    \begin{subfigure}{0.32\textwidth}
        \centering
        \includegraphics[width=\linewidth]{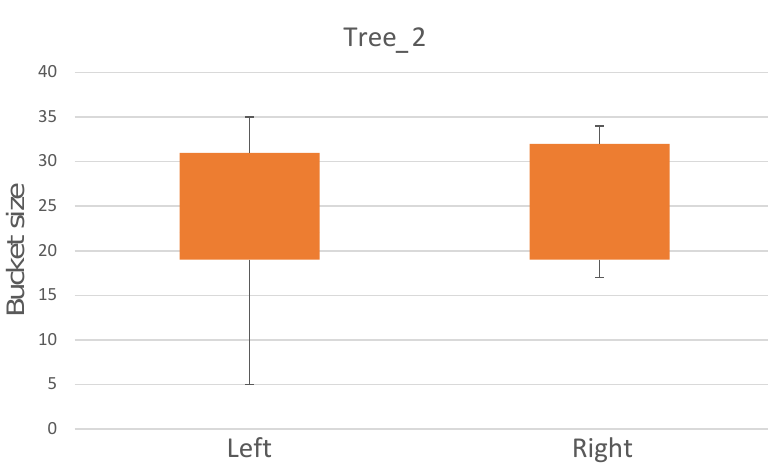}
    \end{subfigure}
    \hfill
    \begin{subfigure}{0.32\textwidth}
        \centering
        \includegraphics[width=\linewidth]{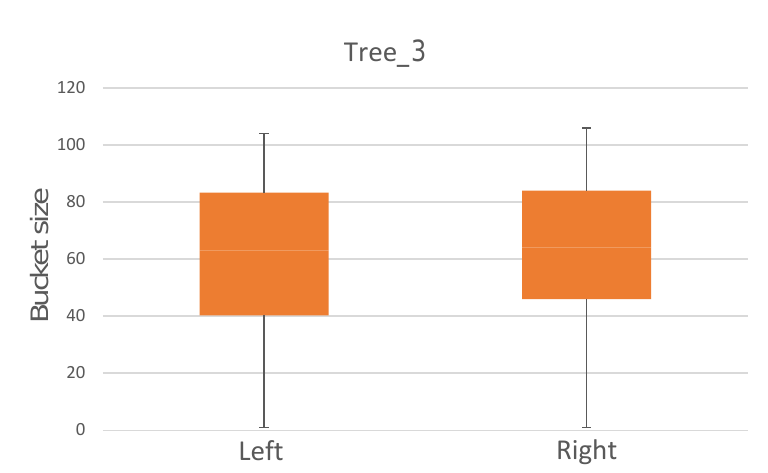}
    \end{subfigure}
    \caption{Bucket data distribution using VBM for Tracking datasets.}
    \label{fig:data_distribution_V_T}
\end{figure}

In the Tracking dataset, both DBM and OBM performed well in maintaining balanced data distributions within the leaf nodes, with bucket sizes peaking around 250 in most trees and median values ranging from 140 to 215, as illustrated in Figures \ref{fig:data_distribution_D_T} and \ref{fig:data_distribution_O_T}. However, a slight imbalance was observed in Tree 5, where bucket sizes were significantly smaller. VBM demonstrated its adaptability by effectively managing varying data volumes, with median bucket sizes ranging from 25 to 60 and peaks nearing 110, as seen in Figure \ref{fig:data_distribution_V_T}, underscoring its effectiveness in optimizing tree structures for datasets of different sizes.

\begin{figure}[H]
    \centering
    \begin{subfigure}{0.49\textwidth}
        \centering
        \includegraphics[width=\linewidth]{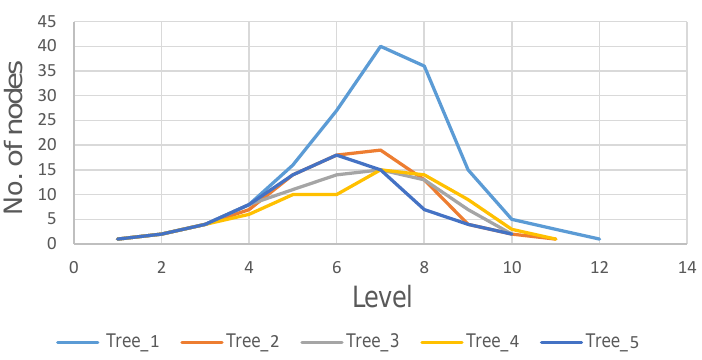}
        \caption{DBM}
        \label{fig:node_distribution_T_D}
    \end{subfigure}
    \hfill
    \begin{subfigure}{0.49\textwidth}
        \centering
        \includegraphics[width=\linewidth]{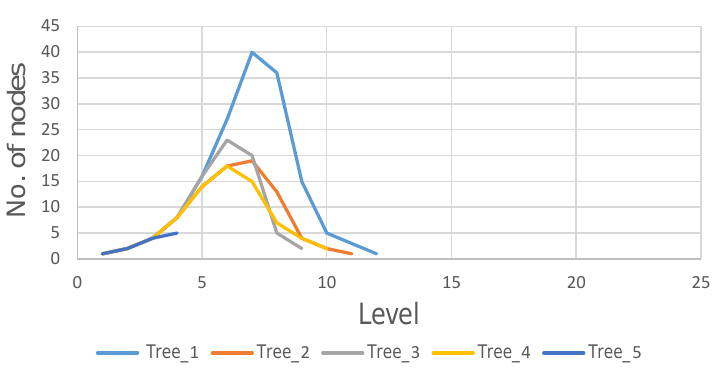}
        \caption{OBM}
        \label{fig:node_distribution_T_O}
    \end{subfigure}
    \hfill
    \begin{subfigure}{0.5\textwidth}
        \centering
        \includegraphics[width=\linewidth]{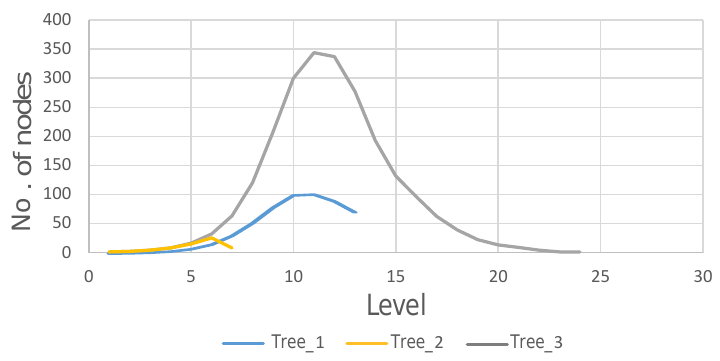}
        \caption{VBM}
        \label{fig:node_distribution_T_V}
    \end{subfigure}
    \caption{Distribution of nodes across tree levels in the Tracking dataset.}
    \label{fig:node_distribution_T}
\end{figure}

The distribution of nodes at each tree level, shown in Figure \ref{fig:node_distribution_T}, further underscores the structural efficiency of these methods. DBM produced trees with a gradual increase in nodes per level, peaking between levels 6 and 8 with a median of 15–20 nodes (Figure \ref{fig:node_distribution_T_D}). The gradual slope suggests that the tree may remain balanced as the data increases. However, in Tree 5, the slope rapidly decreases, indicating an unbalanced tree, as previously observed. OBM exhibited a similar range of levels, from 6 to 12, but with an increased number of nodes ranging from 17 to 22, and peaks reaching up to 40 nodes. This structure suggests deeper trees with concentrated nodes, which may enhance data partitioning but could also increase traversal costs (Figure \ref{fig:node_distribution_T_O}). Unlike DBM, OBM's rapid slope suggests that the tree may become unbalanced as the data grows. In contrast, VBM, shown in Figure \ref{fig:node_distribution_T_V}, displayed a broader node distribution, peaking at around 350 nodes, with levels extending up to 24. The slow slope in VBM indicates its ability to maintain a balanced node distribution across various levels, contributing to a more stable and efficient search process.

\begin{figure}[h]
    \centering
    \begin{subfigure}{0.32\textwidth}
        \centering
        \includegraphics[width=\linewidth]{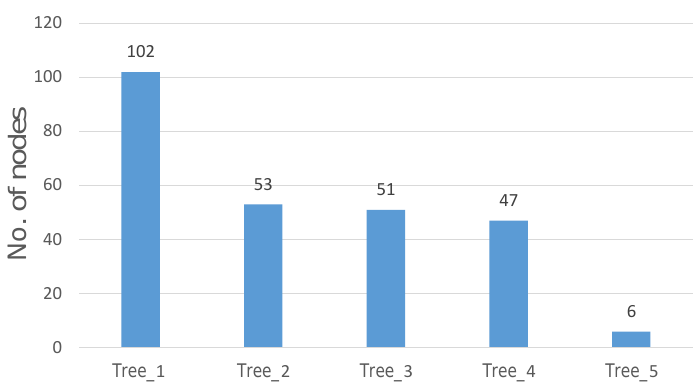}
        \caption{DBM}
        \label{fig:InterNode_D_T}
    \end{subfigure}
    \hfill
    \begin{subfigure}{0.32\textwidth}
        \centering
        \includegraphics[width=\linewidth]{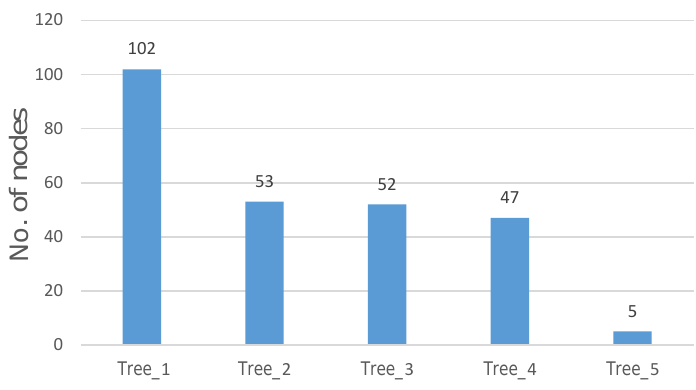}
        \caption{OBM}
        \label{fig:InterNode_O_T}
    \end{subfigure}
    \hfill
    \begin{subfigure}{0.32\textwidth}
        \centering
        \includegraphics[width=\linewidth]{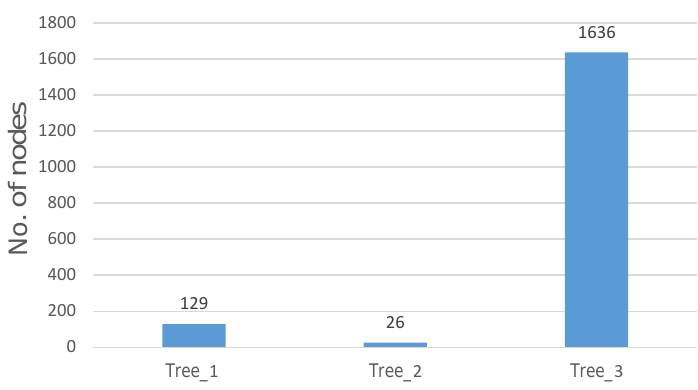}
        \caption{VBM}
        \label{fig:InterNode_V_T}
    \end{subfigure}
    \caption{Number of Internal Nodes in the Tracking dataset.}
    \label{fig:InterNode_T}
\end{figure}

\begin{figure}[h]
    \centering
    \begin{subfigure}{0.32\textwidth}
        \centering
        \includegraphics[width=\linewidth]{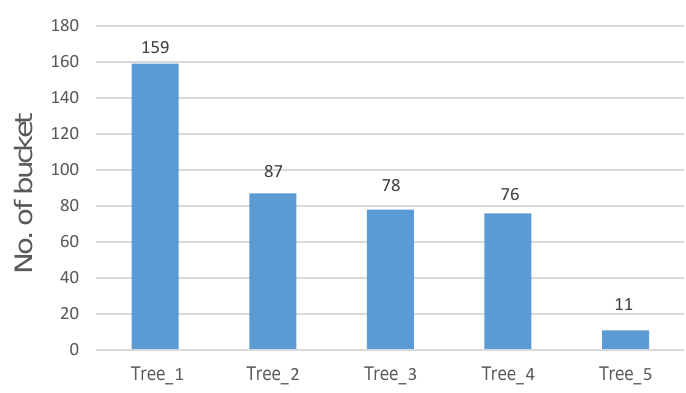}
        \caption{DBM}
        \label{fig:Bucket_D_T}
    \end{subfigure}
    \hfill
    \begin{subfigure}{0.32\textwidth}
        \centering
        \includegraphics[width=\linewidth]{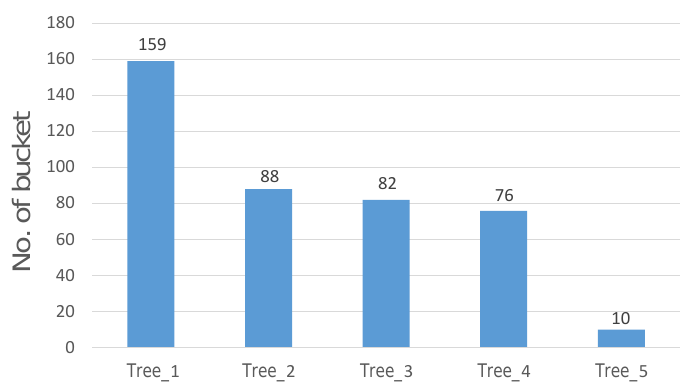}
        \caption{OBM}
        \label{fig:Bucket_O_T}
    \end{subfigure}
    \hfill
    \begin{subfigure}{0.32\textwidth}
        \centering
        \includegraphics[width=\linewidth]{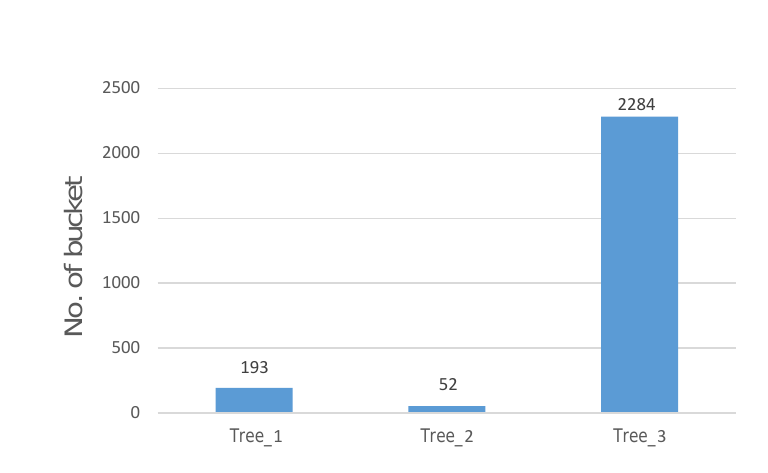}
        \caption{VBM}
        \label{fig:Bucket_V_T}
    \end{subfigure}
    \caption{Number of Nodes Leaves (Buckets) in the Tracking dataset.}
    \label{fig:Bucket_T}
\end{figure}

\begin{figure}[H]
    \centering
    \begin{subfigure}{0.32\textwidth}
        \centering
        \includegraphics[width=\linewidth]{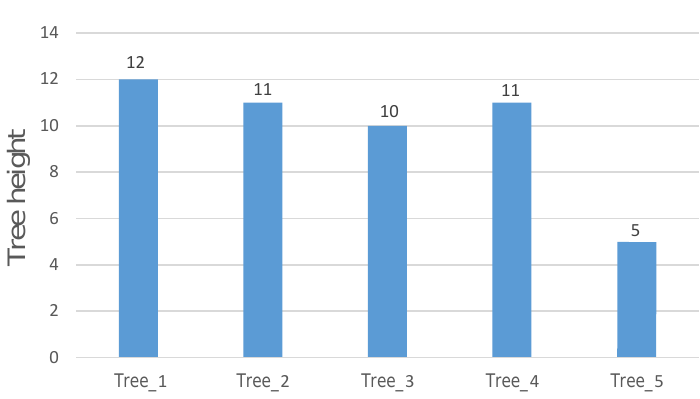}
        \caption{DBM}
        \label{fig:Hauteur_D_T}
    \end{subfigure}
    \hfill
    \begin{subfigure}{0.32\textwidth}
        \centering
        \includegraphics[width=\linewidth]{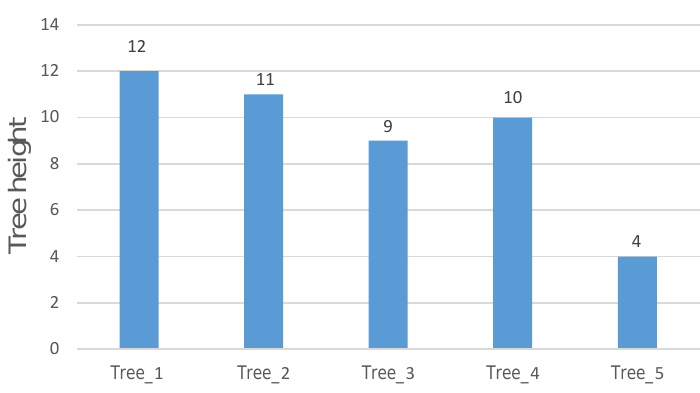}
        \caption{OBM}
        \label{fig:Hauteur_O_T}
    \end{subfigure}
    \hfill
    \begin{subfigure}{0.32\textwidth}
        \centering
        \includegraphics[width=\linewidth]{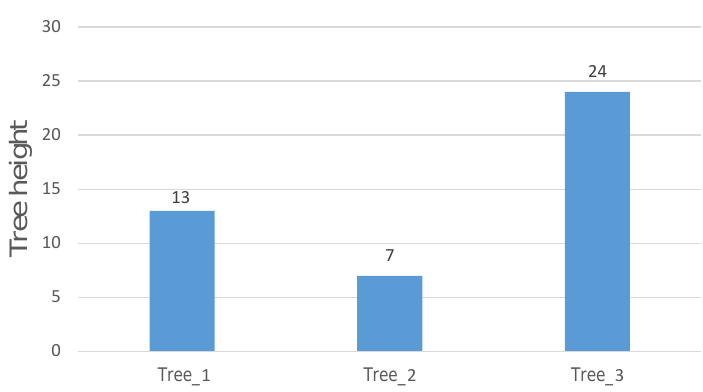}
        \caption{VBM}
        \label{fig:Hauteur_V_T}
    \end{subfigure}
    \caption{Height of the Trees in the Tracking dataset.}
    \label{fig:Hauteur_T}
\end{figure}

Figures \ref{fig:InterNode_T}, \ref{fig:Bucket_T}, and \ref{fig:Hauteur_T} provide additional insights into the internal structure of the trees. Both DBM and OBM consistently produced balanced trees, with internal nodes ranging from 47 to 102, tree heights between 9 and 12, and leaf nodes ranging from 76 to 159. These results suggest that both methods are effective in maintaining moderate tree depth and efficient partitioning, supporting reliable data retrieval. In contrast, VBM exhibited greater variability, with internal nodes ranging from 26 to 1636, tree heights from 7 to 24, and leaf nodes from 52 to 2284. Despite its larger size across these dimensions, VBM demonstrates a well-filled and balanced structure, making it a highly adaptable solution for managing more complex data distributions.

\subsubsection{WARD dataset}

\begin{figure}[h]
    \centering
    \begin{subfigure}{0.32\textwidth}
        \centering
        \includegraphics[width=\linewidth]{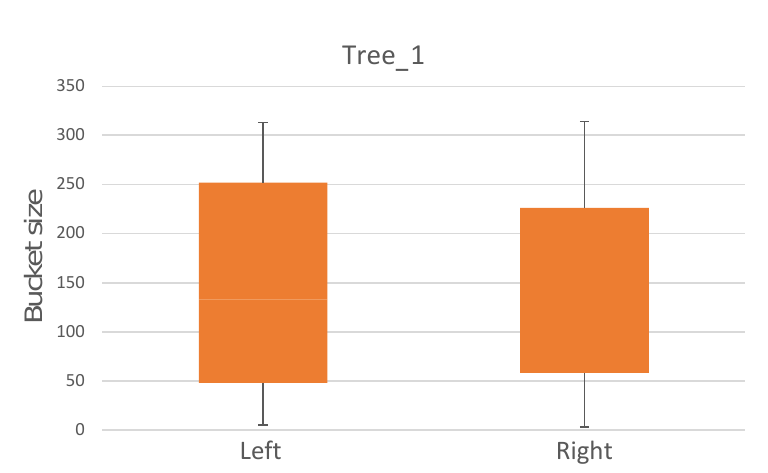}
    \end{subfigure}
    \hfill
    \begin{subfigure}{0.32\textwidth}
        \centering
        \includegraphics[width=\linewidth]{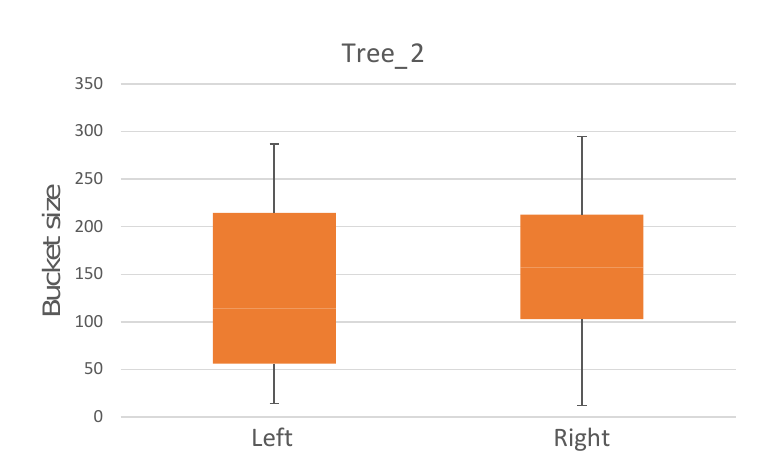}
    \end{subfigure}
    \hfill
    \begin{subfigure}{0.32\textwidth}
        \centering
        \includegraphics[width=\linewidth]{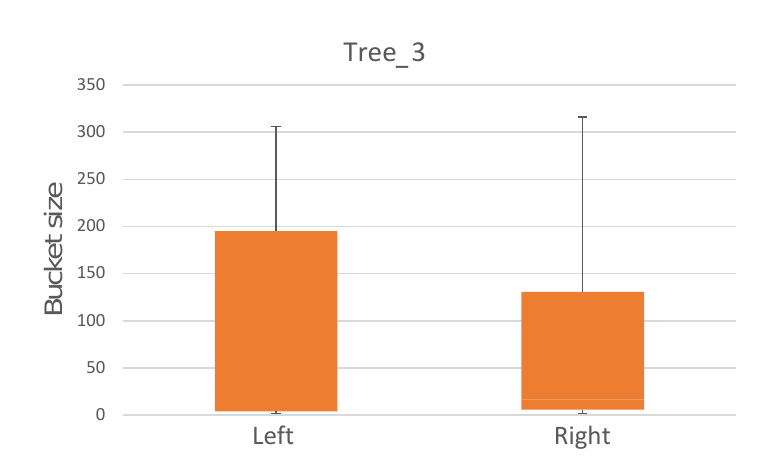}
    \end{subfigure}
    \vspace{1em} 
    \begin{subfigure}{0.32\textwidth}
        \centering
        \includegraphics[width=\linewidth]{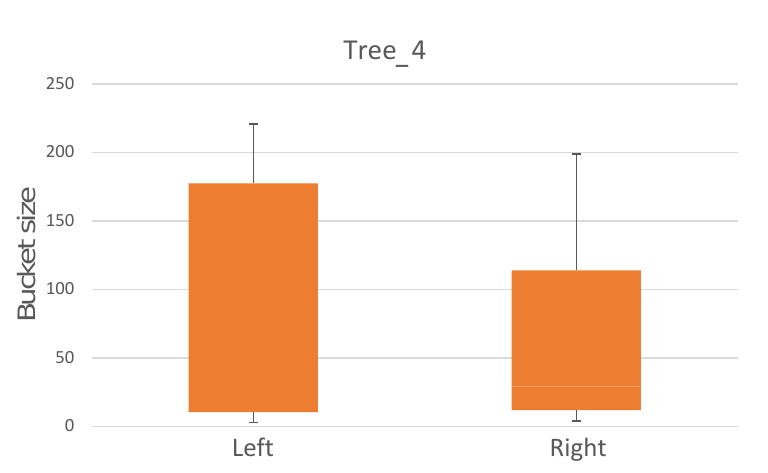}
    \end{subfigure}
    \begin{subfigure}{0.32\textwidth}
        \centering
        \includegraphics[width=\linewidth]{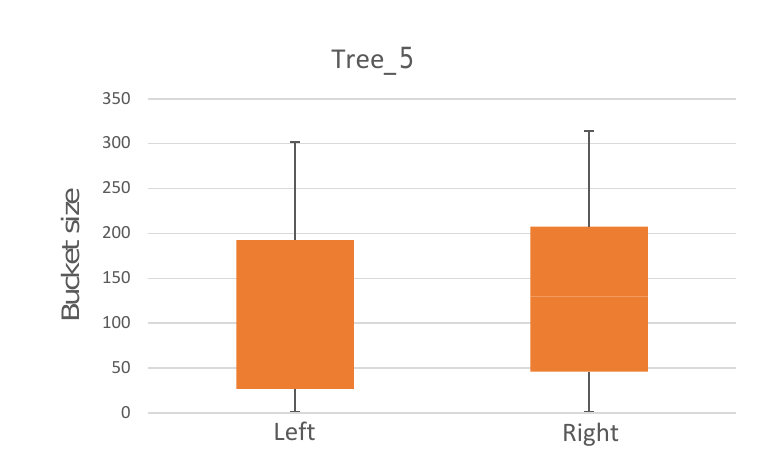}
    \end{subfigure}
        \begin{subfigure}{0.32\textwidth}
        \centering
        \includegraphics[width=\linewidth]{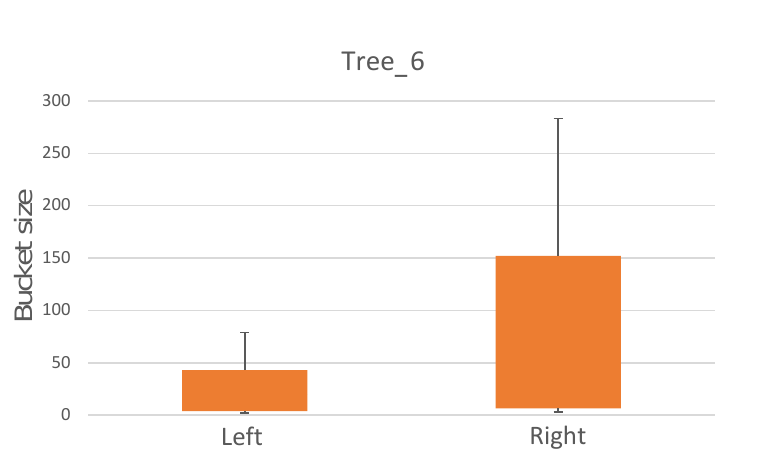}
    \end{subfigure}
    \caption{Bucket data distribution using DBM for WARD datasets.}
    \label{fig:data_distribution_D_W}
\end{figure}

\begin{figure}[h]
    \centering
    \begin{subfigure}{0.4\textwidth}
        \centering
        \includegraphics[width=\linewidth]{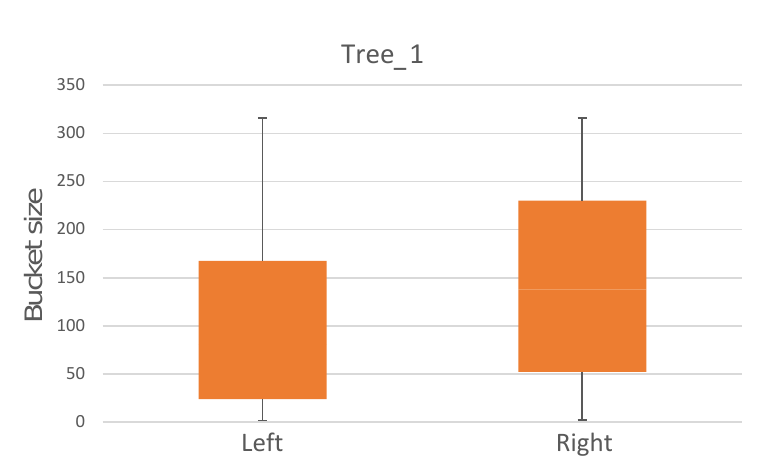}
    \end{subfigure}
    \hspace{3em}
    \begin{subfigure}{0.4\textwidth}
        \centering
        \includegraphics[width=\linewidth]{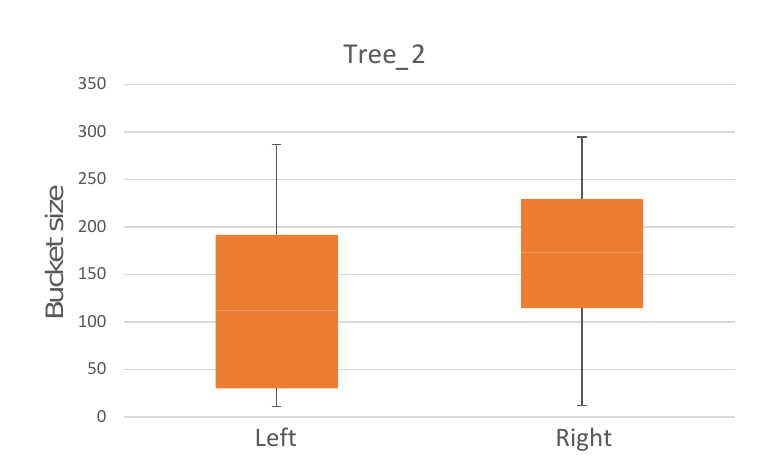}
    \end{subfigure}
    \vspace{1em} 
    \begin{subfigure}{0.4\textwidth}
        \centering
        \includegraphics[width=\linewidth]{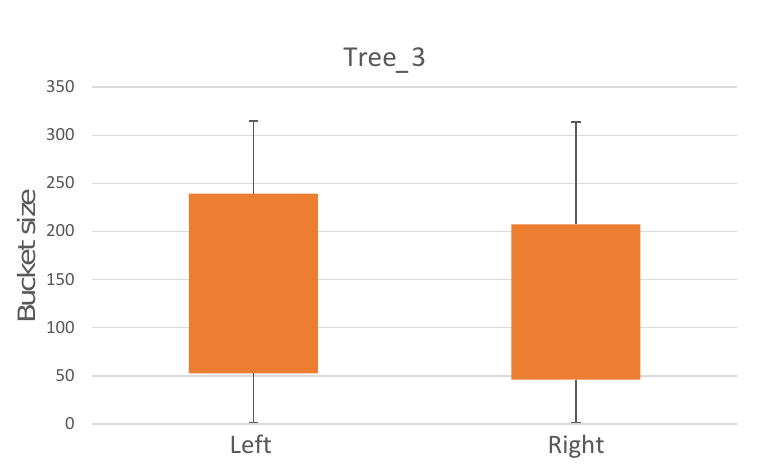}
    \end{subfigure}
    \hspace{3em}
    \begin{subfigure}{0.4\textwidth}
        \centering
        \includegraphics[width=\linewidth]{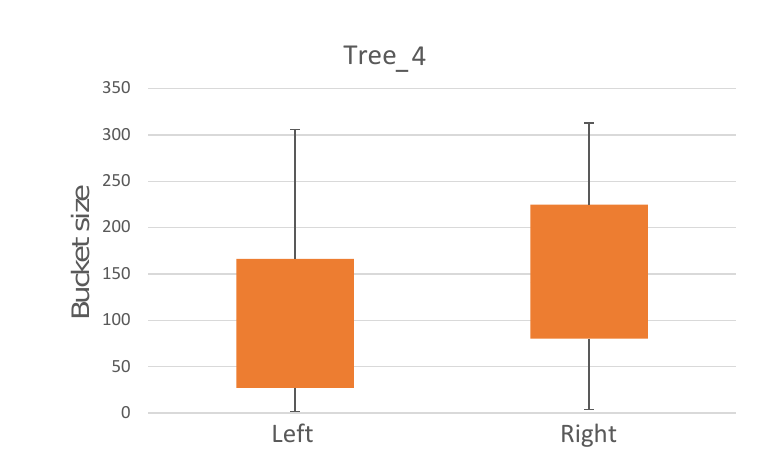}
    \end{subfigure}
    \caption{Bucket data distribution using OBM for WARD datasets.}
    \label{fig:data_distribution_O_W}
\end{figure}

\begin{figure}[h]
    \centering
    \begin{subfigure}{0.32\textwidth}
        \centering
        \includegraphics[width=\linewidth]{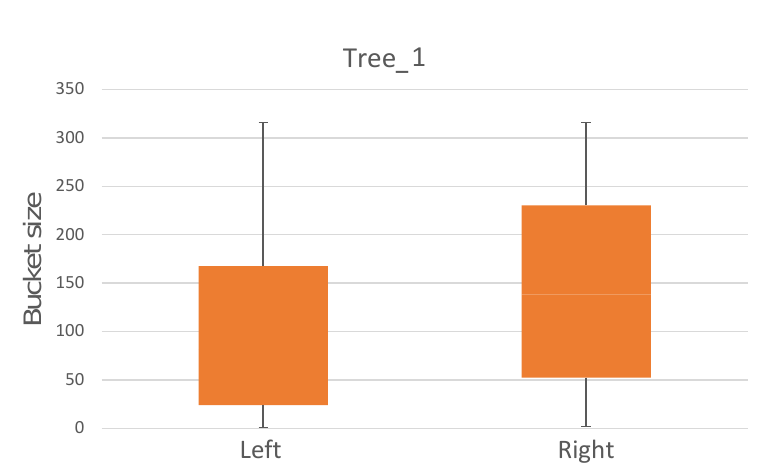}
    \end{subfigure}
    \hfill
    \begin{subfigure}{0.32\textwidth}
        \centering
        \includegraphics[width=\linewidth]{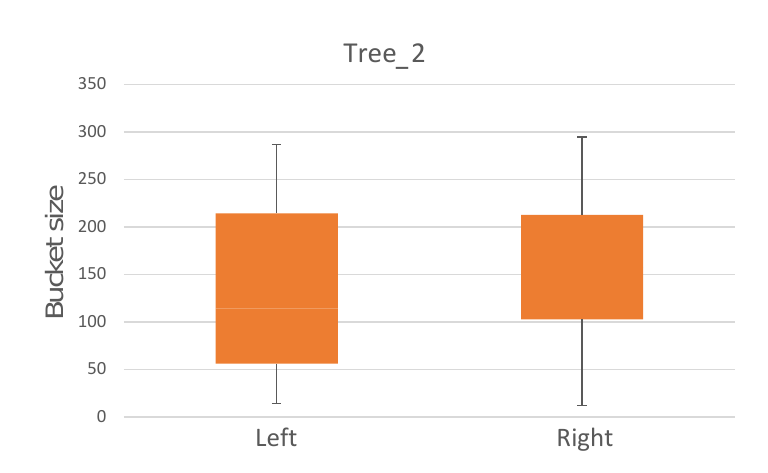}
    \end{subfigure}
    \hfill
    \begin{subfigure}{0.32\textwidth}
        \centering
        \includegraphics[width=\linewidth]{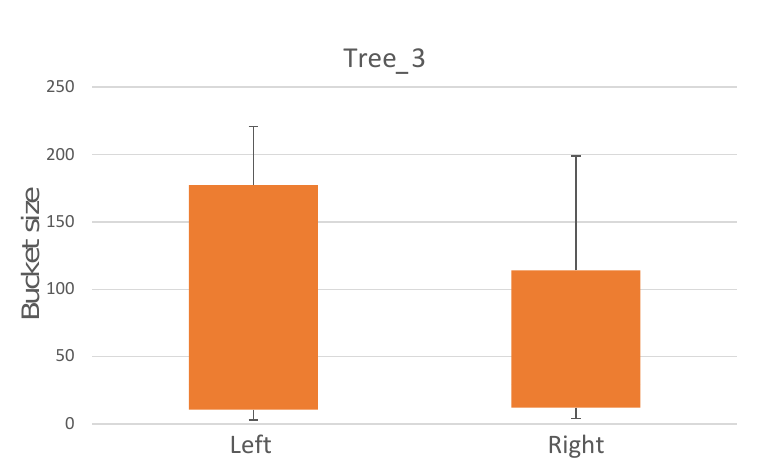}
    \end{subfigure}
    \vspace{1em} 
    \begin{subfigure}{0.32\textwidth}
        \centering
        \includegraphics[width=\linewidth]{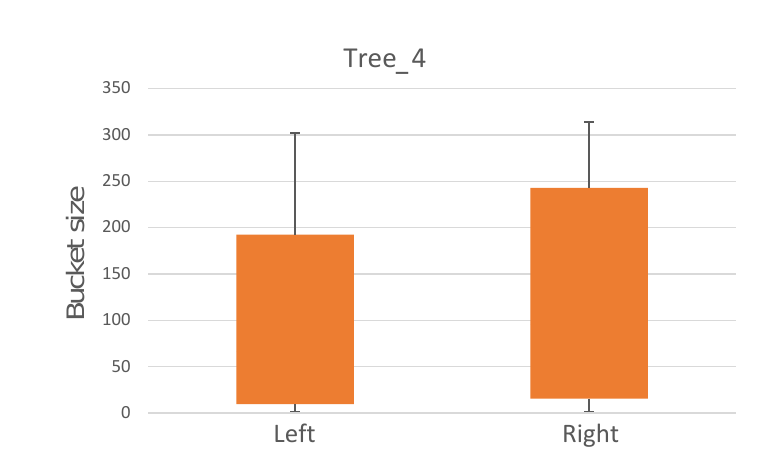}
    \end{subfigure}
    \begin{subfigure}{0.32\textwidth}
        \centering
        \includegraphics[width=\linewidth]{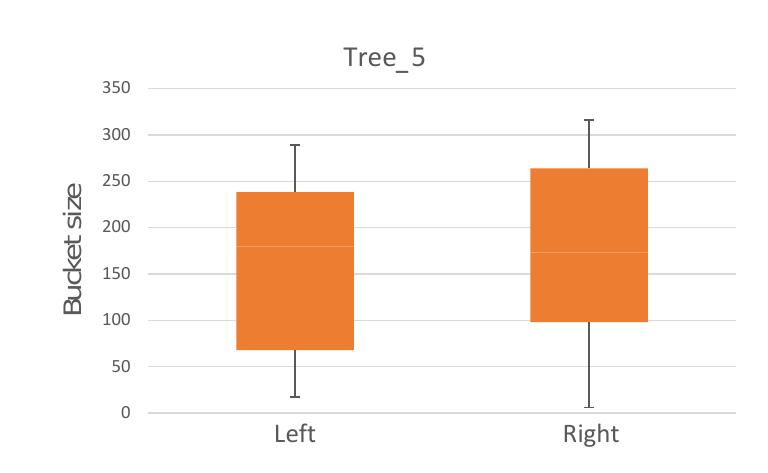}
    \end{subfigure}
        \begin{subfigure}{0.32\textwidth}
        \centering
        \includegraphics[width=\linewidth]{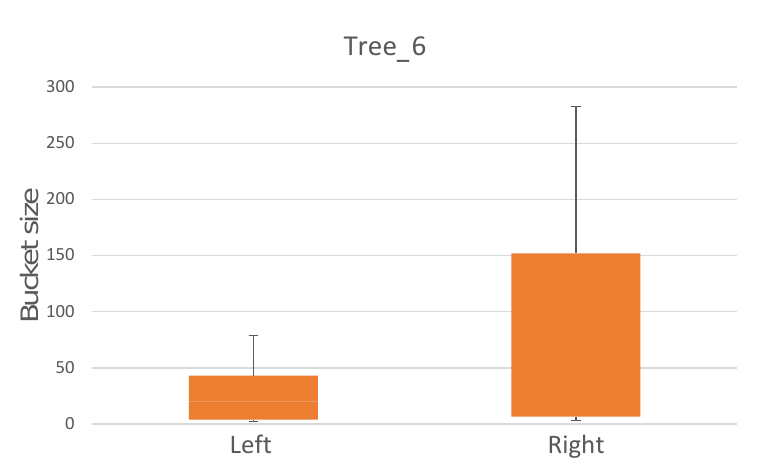}
    \end{subfigure}
    \caption{Bucket data distribution using VBM for WARD datasets.}
    \label{fig:data_distribution_V_W}
\end{figure}

In the WARD dataset, all three methods continued to perform strongly, showing consistent results across various metrics. DBM maintained median bucket sizes around 120 to 160 in most trees (Figure \ref{fig:data_distribution_D_W}), with only a slight reduction observed on the left side of Tree 6. OBM exhibited uniform distributions, as seen in Figure \ref{fig:data_distribution_O_W}, with median bucket sizes also around 120 to 160 and a peak around 230. VBM similarly maintained median bucket sizes around 250, with a slight reduction on the left side of Tree 6 (Figure \ref{fig:data_distribution_V_W}).

\begin{figure}[H]
    \centering
    \begin{subfigure}{0.49\textwidth}
        \centering
        \includegraphics[width=\linewidth]{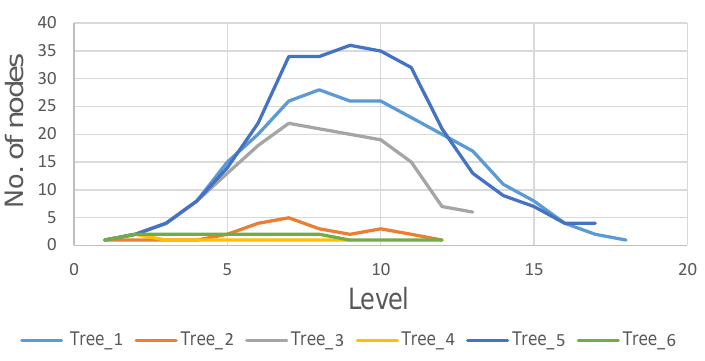}
        \caption{DBM}
        \label{fig:node_distribution_W_D}
    \end{subfigure}
    \hfill
    \begin{subfigure}{0.49\textwidth}
        \centering
        \includegraphics[width=\linewidth]{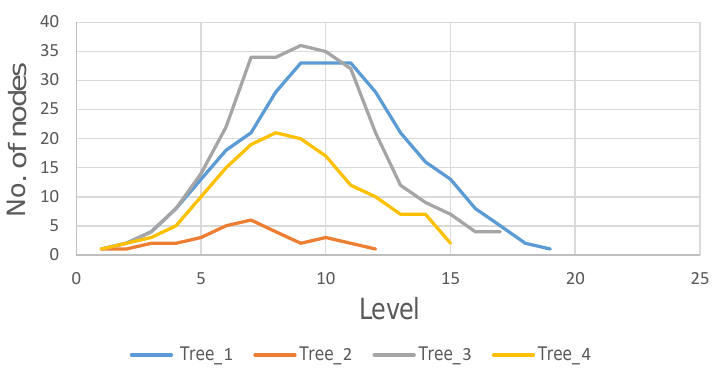}
        \caption{OBM}
        \label{fig:node_distribution_W_O}
    \end{subfigure}
    \hfill
    \begin{subfigure}{0.5\textwidth}
        \centering
        \includegraphics[width=\linewidth]{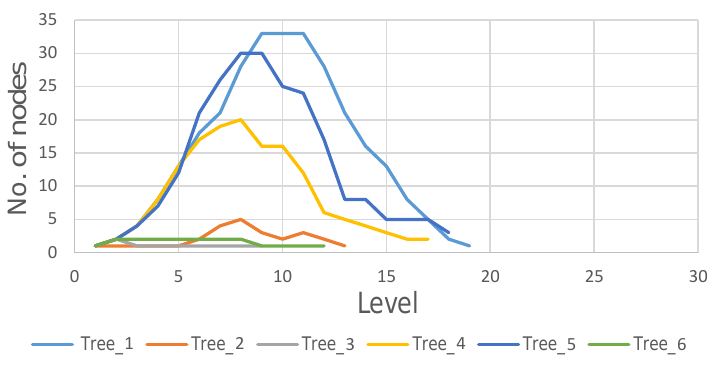}
        \caption{VBM}
        \label{fig:node_distribution_W_V}
    \end{subfigure}
    \caption{Distribution of nodes across tree levels in the WARD dataset.}
    \label{fig:node_distribution_W}
\end{figure}

The results of node distribution across tree levels, as shown in Figure \ref{fig:node_distribution_W}, reveal consistent performance across all three methods—DBM, OBM, and VBM—with node counts peaking between 30 and 36 at levels 10 to 18. The gradual slopes indicate that each method effectively maintains a balanced distribution of nodes, with tree levels being always well-filled, thereby promoting efficient search operations.

\begin{figure}[h]
    \centering
    \begin{subfigure}{0.32\textwidth}
        \centering
        \includegraphics[width=\linewidth]{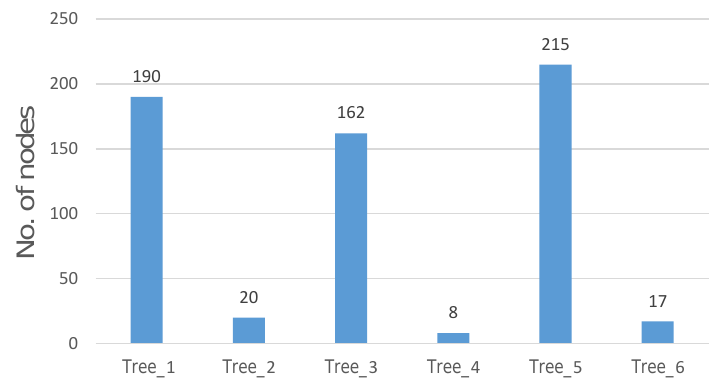}
        \caption{DBM}
        \label{fig:InterNode_D_W}
    \end{subfigure}
    \hfill
    \begin{subfigure}{0.32\textwidth}
        \centering
        \includegraphics[width=\linewidth]{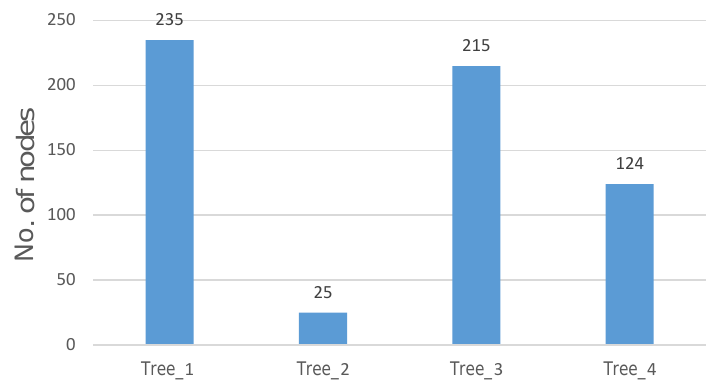}
        \caption{OBM}
        \label{fig:InterNode_O_W}
    \end{subfigure}
    \hfill
    \begin{subfigure}{0.32\textwidth}
        \centering
        \includegraphics[width=\linewidth]{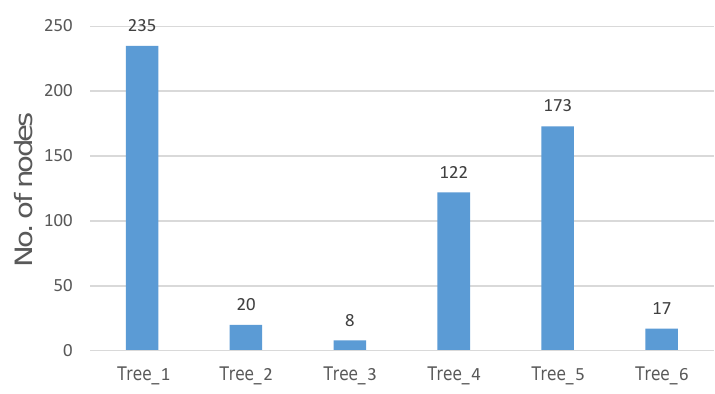}
        \caption{VBM}
        \label{fig:InterNode_V_W}
    \end{subfigure}
    \caption{Number of Internal Nodes in the WARD dataset.}
    \label{fig:InterNode_W}
\end{figure}

\begin{figure}[h]
    \centering
    \begin{subfigure}{0.32\textwidth}
        \centering
        \includegraphics[width=\linewidth]{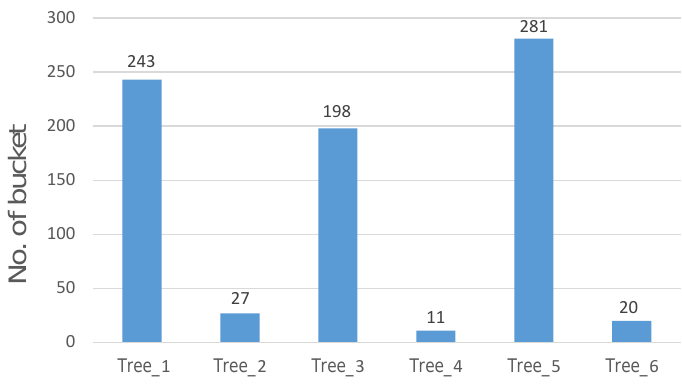}
        \caption{DBM}
        \label{fig:Bucket_D_W}
    \end{subfigure}
    \hfill
    \begin{subfigure}{0.32\textwidth}
        \centering
        \includegraphics[width=\linewidth]{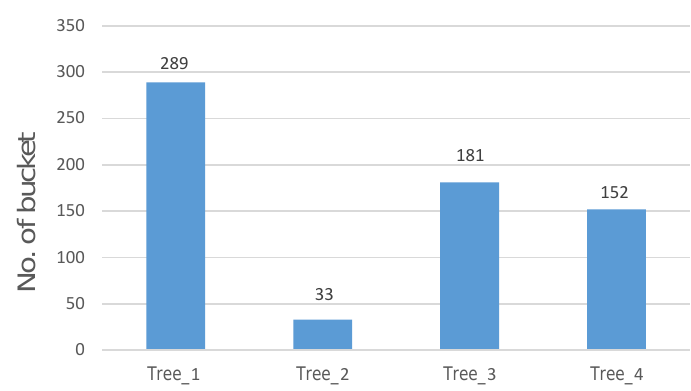}
        \caption{OBM}
        \label{fig:Bucket_O_W}
    \end{subfigure}
    \hfill
    \begin{subfigure}{0.32\textwidth}
        \centering
        \includegraphics[width=\linewidth]{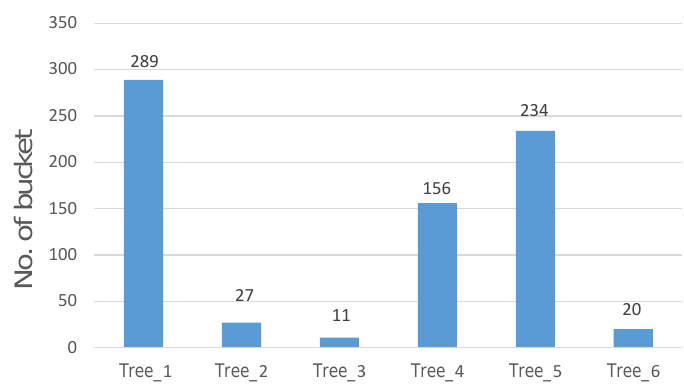}
        \caption{VBM}
        \label{fig:Bucket_V_W}
    \end{subfigure}
    \caption{Number of Nodes Leaves (Buckets) in the WARD dataset.}
    \label{fig:Bucket_W}
\end{figure}

\begin{figure}[h]
    \centering
    \begin{subfigure}{0.32\textwidth}
        \centering
        \includegraphics[width=\linewidth]{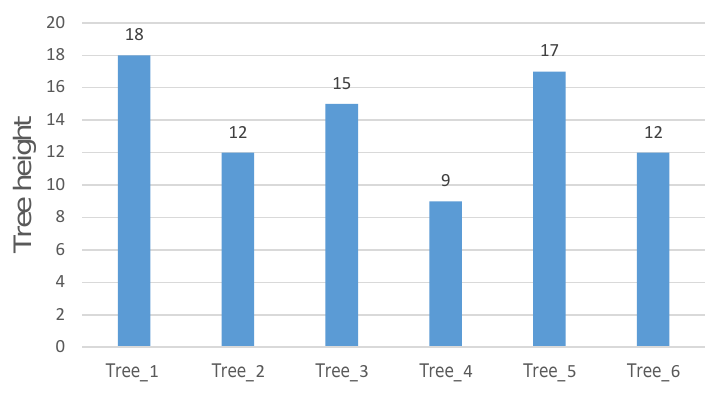}
        \caption{DBM}
        \label{fig:Hauteur_D_W}
    \end{subfigure}
    \hfill
    \begin{subfigure}{0.32\textwidth}
        \centering
        \includegraphics[width=\linewidth]{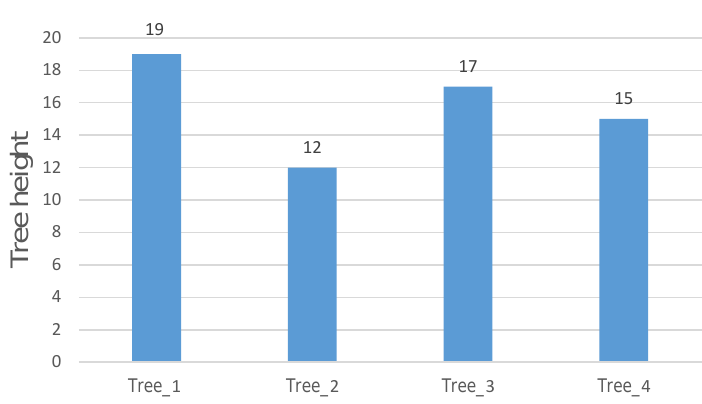}
        \caption{OBM}
        \label{fig:Hauteur_O_W}
    \end{subfigure}
    \hfill
    \begin{subfigure}{0.32\textwidth}
        \centering
        \includegraphics[width=\linewidth]{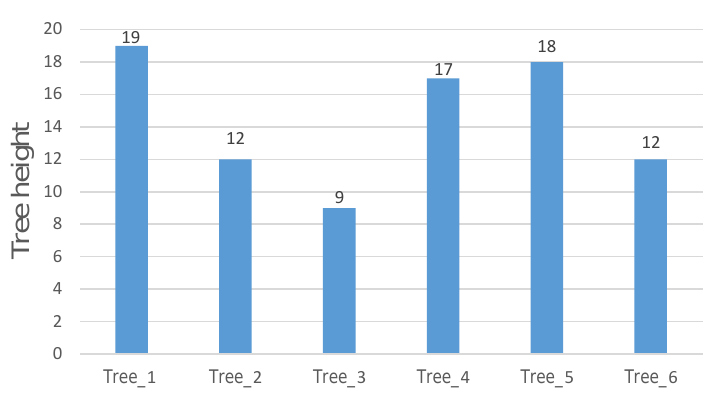}
        \caption{VBM}
        \label{fig:Hauteur_V_W}
    \end{subfigure}
    \caption{Height of the Trees in the WARD dataset.}
    \label{fig:Hauteur_W}
\end{figure}

Finally, Figures \ref{fig:InterNode_W}, \ref{fig:Bucket_W}, and \ref{fig:Hauteur_W} illustrate the number of internal nodes, leaf nodes, and tree heights for the WARD dataset. DBM and OBM continued to demonstrate balanced structures, with internal nodes ranging from 17 to 215, tree heights between 9 and 19, and leaf nodes ranging from 11 to 281. VBM, once again, displayed a broader range, with internal nodes varying from 8 to 235, tree heights from 9 to 19, and leaf nodes from 11 to 289. This variability highlights VBM's flexibility in managing diverse data complexities and its ability to adapt the tree structure to the specific demands of the dataset.

\subsection{Construction Cost}

\begin{figure}[h]
    \centering
    \begin{subfigure}{0.49\textwidth}
        \centering
        \includegraphics[width=\linewidth]{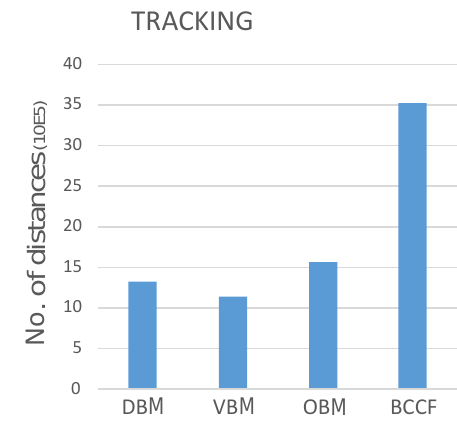}
        \caption{}
        \label{fig:distances_T}
    \end{subfigure}
    \hfill
    \begin{subfigure}{0.49\textwidth}
        \centering
        \includegraphics[width=\linewidth]{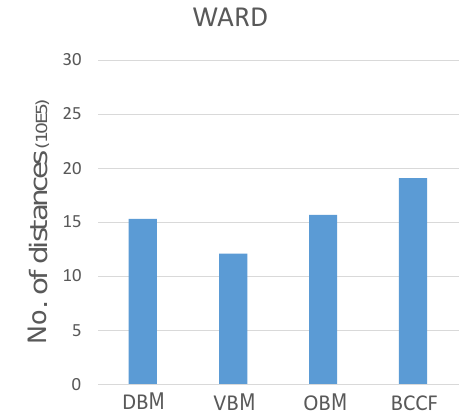}
        \caption{}
        \label{fig:distances_W}
    \end{subfigure}
    \hfill
    \begin{subfigure}{0.49\textwidth}
        \centering
        \includegraphics[width=\linewidth]{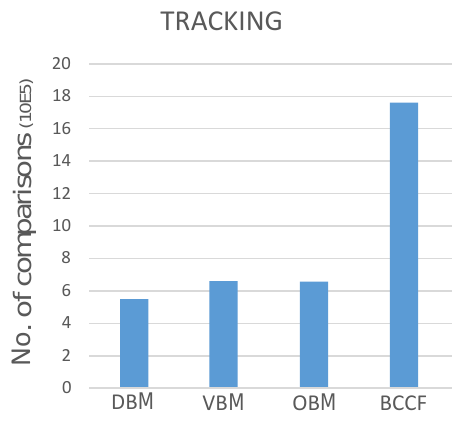}
        \caption{}
        \label{fig:comparisons_T}
    \end{subfigure}
    \hfill
    \begin{subfigure}{0.49\textwidth}
        \centering
        \includegraphics[width=\linewidth]{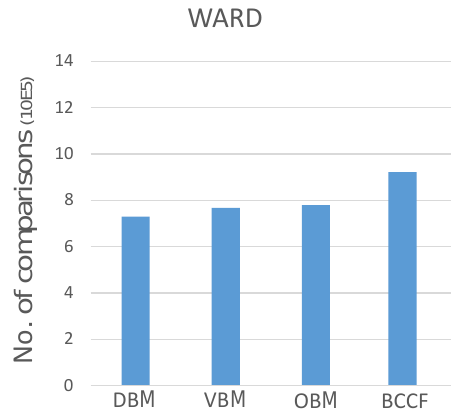}
        \caption{}
        \label{fig:comparisons_W}
    \end{subfigure}
    \caption{Construction cost.}
    \label{fig:Construction_cost}
\end{figure}

To evaluate the performance of the proposed structures compared to the BCCF-tree, we analyzed the construction costs in terms of the number of distances calculated and the number of comparisons made during the construction of indexes. The results, presented in Figure \ref{fig:Construction_cost}, provide a clear comparison of these metrics, with the container size fixed as specified in Table \ref{tab:datasets}.


Across both the Tracking and WARD datasets, the VBM method consistently emerges as the most effective approach. Structural evaluations revealed that VBM produces balanced trees with the flexibility to adapt to varying data complexities, which is directly reflected in its construction costs. In the Tracking dataset, VBM recorded the lowest distance cost at 11.7, compared to 13.6 for DBM and 16.1 for OBM (see Figure \ref{fig:distances_T}), while managing a comparison cost of 6.8, slightly higher than DBM's 5.7 and OBM's 6.5 (see Figure \ref{fig:comparisons_T}). The trends are reaffirmed in the WARD dataset, where VBM again demonstrates its superiority with a lower distance cost of 12.5, compared to 15.7 for DBM and 16.1 for OBM, as shown in Figure \ref{fig:distances_W}. According to Figure \ref{fig:comparisons_W}, the comparison cost of VBM in WARD is slightly higher at 7.8, compared to DBM's 7.4, but it remains competitive. Compared to the BCCF-tree, which recorded much higher costs (36.6 in distance and 17.6 in comparisons), all three methods, particularly VBM, show significant improvements in efficiency across both datasets. This efficiency underscores VBM's ability to maintain a well-distributed structure with minimal overlap.

\subsection{Search Efficiency}

\begin{figure}[h]
    \centering
    \begin{subfigure}{0.49\textwidth}
        \centering
        \includegraphics[width=\linewidth, page=1]{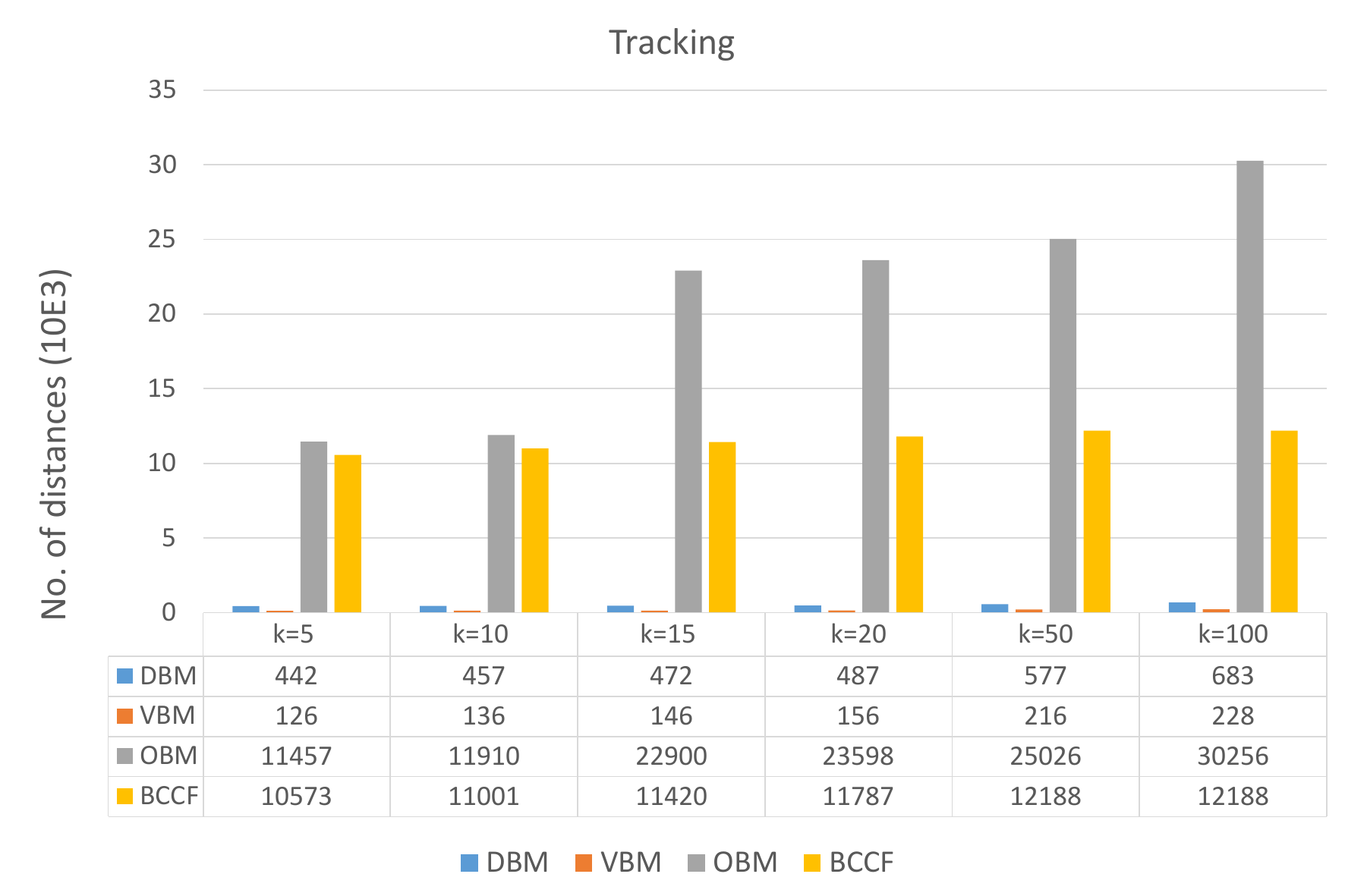}
        \caption{}
        \label{fig:knn_distances_T}
    \end{subfigure}
    \hfill
    \begin{subfigure}{0.49\textwidth}
        \centering
        \includegraphics[width=\linewidth, page=2]{Figs/Results/Knn_Search/Knn_Search_WARD.pdf}
        \caption{}
        \label{fig:knn_distances_W}
    \end{subfigure}
    \hfill
    \begin{subfigure}{0.49\textwidth}
        \centering
        \includegraphics[width=\linewidth, page=2]{Figs/Results/Knn_Search/Knn_Search_Tracking.pdf}
        \caption{}
        \label{fig:knn_comparison_T}
    \end{subfigure}
    \hfill
    \begin{subfigure}{0.49\textwidth}
        \centering
        \includegraphics[width=\linewidth, page=1]{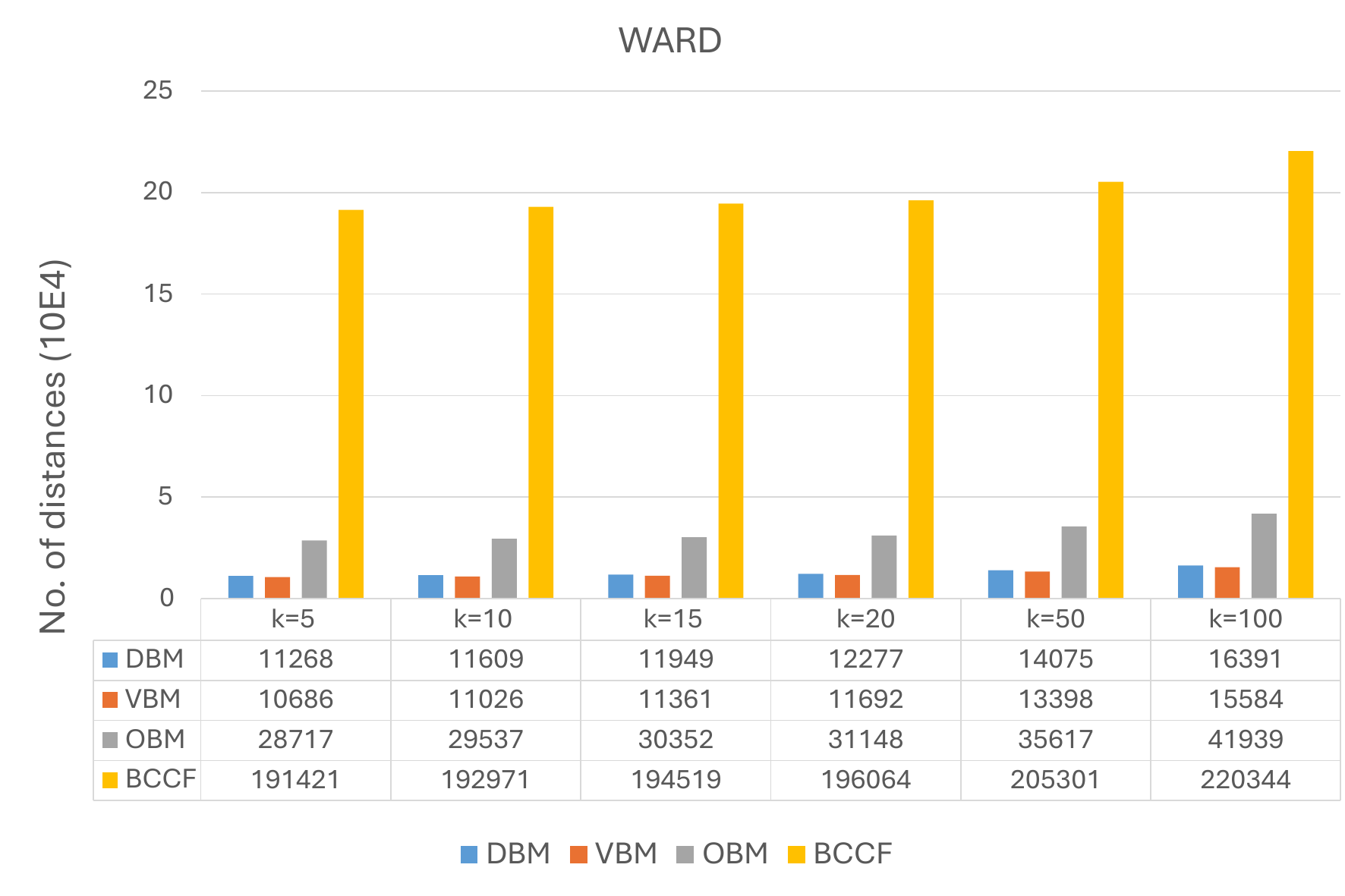}
        \caption{}
        \label{fig:knn_comparison_W}
    \end{subfigure}
    \begin{subfigure}{0.49\textwidth}
        \centering
        \includegraphics[width=\linewidth, page=3]{Figs/Results/Knn_Search/Knn_Search_Tracking.pdf}
        \caption{}
        \label{fig:knn_Time_T}
    \end{subfigure}
    \hfill
    \begin{subfigure}{0.49\textwidth}
        \centering
        \includegraphics[width=\linewidth, page=3]{Figs/Results/Knn_Search/Knn_Search_WARD.pdf}
        \caption{}
        \label{fig:knn_Time_W}
    \end{subfigure}

    \caption{Performance of the similarity search algorithm.}
    \label{fig:knn_Search}
\end{figure}

The similarity search algorithm's performance was evaluated against the BCCF-tree structure by analyzing the average number of distance calculations, comparisons performed, and execution times for fulfilling 100 $k$NN requests across different $k$ values ($k$=5, 10, 15, 20, 50, 100). The results of these experiments are presented in Figure \ref{fig:knn_Search}.

In the Tracking dataset, VBM consistently demonstrates superior performance across all metrics. At \( k = 5 \), VBM records just 126 distance calculations, significantly lower than DBM's 442 and vastly outperforming OBM's 11,457 and BCCF's 10,573 (see Figure \ref{fig:knn_distances_T}). This trend continues as \( k \) increases, with VBM consistently maintaining the lowest distance counts, highlighting its efficiency in managing proximity calculations during searches. In terms of comparisons (see Figure \ref{fig:knn_comparison_T}), VBM again outperforms the other methods, recording only 532 comparisons at \( k = 5 \), compared to DBM's 1,997, OBM's 13,696, and BCCF's 16,447. This efficiency is further reflected in the search time, where VBM consistently requires the least time across all values of \( k \). For example, at \( k = 100 \), VBM completes the search in just 0.02 seconds, significantly faster than DBM's 0.045 seconds and OBM's 0.07 seconds, while BCCF lags considerably at 1.2 seconds (see Figure \ref{fig:knn_Time_T}).

Similarly, in the WARD dataset, VBM continues to outperform the other methods. At \( k = 5 \), VBM records 10,686 distance calculations, slightly lower than DBM's 11,268 and significantly better than OBM's 28,717 and BCCF's staggering 191,421 (see Figure \ref{fig:knn_distances_W}). This pattern persists across all \( k \) values, with VBM consistently maintaining lower distance and comparison counts, demonstrating its efficiency in processing large datasets. In terms of comparisons, VBM remains strong, with 47,147 comparisons at \( k = 5 \), compared to DBM's 50,660, OBM's 129,011, and BCCF's 922,127 (see Figure \ref{fig:knn_comparison_W}). This superior performance in both distance and comparison metrics underscores VBM's ability to handle complex data efficiently. Regarding execution time, VBM continues to be competitive, often matching or slightly exceeding DBM's performance, while still vastly outperforming OBM and BCCF, especially as \( k \) increases (see Figure \ref{fig:knn_Time_W}).

In the final analysis, VBM emerges as the most efficient method for $k$-NN searches, offering an optimal balance between distance calculations, comparisons, and execution time across both datasets. This superiority is rooted in VBM’s advanced approach to overlap management, which minimizes unnecessary computations by precisely calculating intersection volumes, thereby ensuring that search operations remain focused and efficient. DBM, while slightly trailing VBM, demonstrates strong performance with its proximity-based partitioning, which strikes a commendable balance between computational simplicity and effectiveness, though it incurs marginally higher costs. On the other hand, OBM typically incurs increased computational overhead, making it less suitable for large-scale searches. The baseline BCCF-tree consistently underperforms, underscoring the significant advancements and optimizations introduced by the proposed methods, particularly the VBM, in enhancing $k$-NN search efficiency.

\section{Conclusion}\label{sec6}

This paper has addressed the critical challenge of optimizing tree-based indexing structures in the context of massive IoT data. The overlap in data space partitions during index creation has been a significant obstacle, leading to inefficiencies in data retrieval and increased computational costs. To overcome this, we introduced three innovative heuristics: the Volume-Based Method (VBM), the Distance-Based Method (DBM), and the Object-Based Method (OBM), each designed to strategically minimize overlap and enhance the overall performance of the indexing process. Through extensive simulations and comparative analyses, we demonstrated that the VBM consistently outperforms both DBM and OBM, particularly in terms of structural balance, search efficiency, and construction costs. VBM's ability to maintain a well-distributed and adaptable tree structure while minimizing overlap was evident across both the Tracking and WARD datasets, making it the most effective method among the three. DBM followed closely, providing robust performance with slightly higher costs, while OBM, although effective in specific scenarios, generally incurred higher computational overhead due to its detailed partitioning strategy.

Moreover, in future work, we aim to further enhance the system’s adaptability by exploring dynamic adjustments of overlap thresholds ($\xi_{min}$ and $\xi_{min}$) tailored to the context and specific characteristics of the data. Additionally, we plan to introduce traceability mechanisms to ensure reliable data monitoring and management across different index structures, optimizing the search space and facilitating more efficient parallel search operations.

\backmatter


\bibliography{sn-bibliography}

\end{document}